\documentstyle[11pt,aaspp4]{article}
\def\ltsima{$\; \buildrel < \over \sim \;$}
\def\simlt{\lower.5ex\hbox{\ltsima}}
\def\gtsima{$\; \buildrel > \over \sim \;$}
\def\simgt{\lower.5ex\hbox{\gtsima}}
\def\kms{{\rm\,km\,s^{-1}}}
\def\kpc{{\rm\,kpc}}

\def\msun{{\rm\,M_\odot}}

\def\pc{{\rm\,pc}}

\def\deg{^\circ}

\def\s{\ifmmode \widetilde \else \~\fi}
\def\={\overline}

\def\spose#1{\hbox to 0pt{#1\hss}}

\def\etal{{\it et al.\ }}
\def\cf{{\it cf.\ }}
\def\eg{{\it e.g.,\ }}
\def\ie{{\it i.e.,\ }}
\def\lta{\mathrel{\spose{\lower 3pt\hbox{$\mathchar"218$}}
     \raise 2.0pt\hbox{$\mathchar"13C$}}}
\def\gta{\mathrel{\spose{\lower 3pt\hbox{$\mathchar"218$}}
     \raise 2.0pt\hbox{$\mathchar"13E$}}}
\def\Dt{\spose{\raise 1.5ex\hbox{\hskip3pt$\mathchar"201$}}}	
\def\dt{\spose{\raise 1.0ex\hbox{\hskip2pt$\mathchar"201$}}}	

\def\=={\equiv}

\def\dotsfill{\leaders\hbox to 1em{\hss.\hss}\hfill}

\def\des{Hatzidimitriou\ }
\def\Sgr{The Sagittarius dwarf spheroidal}
\def\sgr{the Sagittarius dwarf spheroidal}
\def\sgg{the Sagittarius dwarf}
\def\Sgg{The Sagittarius dwarf}
\def\IGI{IGI-I}
\def\IGItwo{IGI-II}
\def\Gyr{{\rm\,Gyr}}
\def\kmsd{{\rm\,km/s/degree}}

\slugcomment{Accepted by The Astronomical Journal}

\lefthead{Ibata et al.}
\righthead{Kinematics, Orbit and Survival of the Sgr dSph}

\begin{document}

\title{The Kinematics, Orbit, and Survival\\
of the Sagittarius Dwarf Spheroidal Galaxy}

\author{Rodrigo A. Ibata}
\affil{Department of Physics and Astronomy, UBC; ibata@astro.ubc.ca}

\author{Rosemary F.G. Wyse\altaffilmark{1},\altaffilmark{2}}
\affil{Department of Physics and Astronomy, JHU; wyse@pha.jhu.edu}

\author{Gerard Gilmore\altaffilmark{3}}
\affil{Institute of Astronomy, Madingley Rd. Cambridge CB3 0HA,
UK; gil@ast.cam.ac.uk}

\author{Michael J. Irwin}
\affil{Royal Greenwich Observatory, Madingley Rd, Cambridge CB3 0EZ;
mike@ast.cam.ac.uk}

\author{Nicholas B. Suntzeff}
\affil{Cerro Tololo InterAmerican Observatory\altaffilmark{4},
Casilla 603, La Serena, Chile; nick@ctio.noao.edu}


\altaffiltext{1}{Institute of Astronomy, Madingley Rd. Cambridge CB3 0HA,
UK}

\altaffiltext{2}{Center for Particle Astrophysics, University of California, Berkeley, CA 95124}

\altaffiltext{3}{Institut d'Astrophysique, 98bis Boulevard Arago, 75014 Paris,
France}

\altaffiltext{4}{Cerro Tololo Inter-American Observatory 
is operated by AURA, Inc.\ under contract to the National Science
Foundation.}


\begin{abstract}
\Sgr\ galaxy,  the closest satellite galaxy of  the Milky  Way, has survived
for many orbits about the Galaxy. Extant numerical calculations modeled this
galaxy as a system with a centrally-concentrated mass profile, following the
light, and  found that it should  lose more than  one-half of its mass every
2--4 orbits  and be completely disrupted  long before now.  Apparently \sgr,
and by implication other dSph galaxies, do not have a centrally-concentrated
profile for their dark matter. We develop a model in which  the stars of the
Sgr dwarf are embedded in a  constant-density dark matter halo, representing
the core of a tidally-limited system, and  show that this is consistent with
its survival.  We  present  new photometric  and kinematic observations   of
\sgr\  and show these  data are  consistent  with  this explanation  for the
continued existence of this galaxy. \Sgg\  is being tidally distorted and is
tidally limited,  but is not  disrupted as  yet.  The  corresponding minimum
total mass  is $10^9 \msun$,  while the central mass  to  visual light ratio
$\sim 50$ in Solar units.

Our new photographic photometry  allows the detection of main-sequence stars
of \sgg\ over an area of $22\deg \times 8\deg$. \Sgg\  is prolate, with axis
ratios $\sim$~3:1:1.  For an adopted  distance of $16  \pm  2 \kpc$ from the
Galactic center on the  opposite side of  the  Galaxy to the Sun,  the major
axis is $\gta 9 \kpc$ long and is  aligned approximately normal to the plane
of   the   Milky   Way Galaxy,   roughly   following    the  coordinate line
$\ell=5^\circ$.
 
The central velocity dispersion of giant stars which are members of \sgg\ is
$11.4 \pm 0.7 \kms$ and is consistent  with being constant  over the face of
the galaxy. The gradient in  mean line-of-sight velocity with position along
the major axis, $dv/db$, is $\sim 0 \kms$/degree  in the central regions and
increases in  amplitude to $dv/db=-3 \kms$/degree  over the  outermost three
degrees for which we have data. A first  measurement of the proper motion of
Sgr determines  the component of  its space  velocity parallel  to its major
axis to be $250 \pm 90 \kms$, directed towards the  Galactic Plane. We model
these kinematic data to determine the orbit of \sgg.  Our best fit model has
an  orbital period of  $\lta 1 \Gyr$  and has \sgr\ close to perigalacticon.
This period is shorter, by  about a factor of $\simgt  10$, than the age  of
the bulk of its stellar population.

\end{abstract}




%

\section{Introduction}

The serendipitous discovery of \sgr\ \markcite{me94}(Ibata, Gilmore \& Irwin
1994, \markcite{me95c}1995;  hereafter IGI-I and IGI-II respectively) during
the  course of  a   spectroscopic study  of   the  bulge of the  Milky   Way
\markcite{me95a}(Ibata \& Gilmore  1995a, \markcite{me95b}1995b) provides an
unprecedented  opportunity for detailed study  of  the interaction between a
normal   large disk galaxy  and  one of its satellite   galaxies, and of the
internal structure of a dwarf spheroidal galaxy.

At least some of the dwarf spheroidal companions  of the Milky Way have long
been    suspected  to  contain   large  quantities    of  dense  dark matter
\markcite{fab83}(\eg Faber \& Lin 1983; \markcite{irw95}Irwin \& \des 1995),
which has important implications for the nature of  the constituents of dark
halos.   Most of the dwarf  spheroidals contain stars  with a broad range of
ages and metallicities, which is  unexpected in the simplest explanation for
their low mean metallicities  --- that chemical  evolution was  truncated by
supernovae-driven      winds        \markcite{san65}(\eg     Sandage   1965;
\markcite{dek86}Dekel \&  Silk  1986).  Further,  in the   currently-popular
hierarchical  clustering   picture   of  structure   formation,    such   as
Cold-Dark-Matter   dominated cosmologies,  very   significant  accretion and
merging of smaller systems  occurs during the evolution  of a normal  galaxy
like the Milky Way; is this on-going?

This paper presents   new photometric and  kinematic data  that are  used to
constrain  models of the  orbit  of \sgr. We  propose  a new  model for  the
internal  structure of   \sgr,  consistent  with  these data   and with  the
previously apparently-surprising  ability of the  Sgr dwarf to have survived
many perigalactic passages.   This  model also has  significant implications
for the nature of dark matter and the evolution of dwarf galaxies.

\section{Physical Properties of the Sagittarius Dwarf Spheroidal Galaxy}

\subsection{Photometric Determination of the Spatial Extent}

The  first survey  to map \sgg\   (\IGI) utilized the red horizontal  branch
(RHB)  and clump giant stars which  are members.   The `background noise' in
this case is   a combination  of small   number statistics  and   foreground
Galactic stars with similar  apparent  magnitudes and colors.  The  isopleth
map derived by \IGI\ and  \IGItwo\ (superimposed on Figure~1 below) revealed
that \sgg\ is visible over an approximately $10^\circ \times 4^\circ$ region
centered near  $\ell=5.6^\circ,  b=-14.0^\circ$,  at  the position  of   the
globular cluster M54 (NGC 6715).  However, the intrinsic rarity of RHB stars
limited  this detection to relatively  high density regions and, as \IGItwo\
emphasized, their outermost isopleth had detected only  $\simlt 50\%$ of the
stellar population  of \sgg.   \IGI\  additionally noted that four  globular
clusters were probable members of this galaxy, even though three of them lay
well outside the available isopleths, further supporting an expectation that
those  isopleths  were a  lower limit  on  the  detectable  size of  the Sgr
dwarf. This expectation has been supported in the interim by several studies
which have identified members of \sgg\ along lines-of-sight well outside the
original detection limits.

Deep CCD color magnitude diagrams of the line  of sight towards the globular
cluster   M55  \markcite{fahl96}(Fahlman  \etal  1996, \markcite{mat96}Mateo
\etal 1996) show, in  addition  to the expected   M55 and Galactic stars,  a
well-defined main  sequence and three  RR~Lyrae stars, which have magnitudes
and   colors expected  of  members  of   \sgr.  This  field  is  located  at
$(\ell=8.8^\circ,b=-23.0^\circ)$, approximately  $3^\circ$ away    from  the
outermost contour reported    in \IGI. Furthermore,   a  study  by  the  DUO
microlensing team \markcite{al96}(Alard 1996) in  a $5^\circ \times 5^\circ$
field centered  at $(\ell=3^\circ,  b=-7^\circ)$  has detected  313 RR~Lyrae
stars which have  distance moduli which show them  to be members  of the Sgr
dwarf. The surface density of  these RR~Lyrae stars  increases away from the
Galactic  plane and is a  broad, flat distribution: compelling evidence that
what   is  seen   is  not   clumpy  tidal   debris  (as  was   suggested  by
\markcite{mat96}Mateo \etal  1996 from  their M55 observations),  but simply
the continuation of the isodensity contours  found by \IGI. In the direction
towards  the Galactic    plane,  these RR~Lyrae  stars   are  found   up  to
$b=-4^\circ$, more than $10^\circ$ away from the center defined by \IGI.

At low Galactic latitudes in the direction of the Galactic Bulge, the number
density and strong gradient  of Galactic  stars preclude using  photographic
survey plates to probe down  to the main  sequence turnoff of \sgg. However,
further out along its  major axis in  the  direction away from  the Galactic
plane, it is possible to probe to the full depth of the survey plates. Stars
near the main sequence   turnoff in \sgg\  are much  more numerous  than red
clump stars and they   also  occupy a region   of  the HR diagram   which is
significantly less  contaminated  by  foreground Galactic  stars;  these two
effects make it possible to obtain a  more sensitive detection of the extent
of \sgg\ using main sequence stars than was  obtained in the original survey
of  red clump stars. Isopleth maps  derived from the excess of main-sequence
stars which are members of \sgg\ on  four UKST fields East  of the center of
\sgg\ (IIIaJ sky  survey plates) are shown in  Figure~1, superimposed on the
original isopleth map from \IGI.

The main  sequence  turnoff of  Sgr occurs  at  a B$_{\rm  J}$  magnitude of
$\approx 21.5$, roughly a magnitude above the plate limit  of UKST IIIaJ sky
survey  plates.    Although significant   geometrical vignetting  is present
beyond a radius of 2.7~degrees, the 5~degree separation of the survey fields
--- giving a  1~degree overlap between plates ---  ensures that the majority
of the region used to construct the overall  isopleth map is only marginally
affected by vignetting.

As for the    red clump  isopleth    maps, suitable comparison  regions   of
color-magnitude  space had  to be  defined in order  to remove  the varying,
mainly  foreground, Galactic  contribution.   Unfortunately,  the  R  survey
plates available for    the fields of  interest  do  not go deep   enough to
realistically sample the Sgr main sequence  turnoff region.  This restricted
selection of  comparison regions in  color-magnitude space to limits  on the
$B_{\rm J}$ magnitude only.   Two samples were generated  for each plate:  a
comparison region in  the magnitude range 19.5 --  21.0; and a main sequence
reference region from 21.5 -- 22.5.  Clearly, there is some contamination in
the comparison  sequence  from  Sgr members  but it  is   a relatively small
fraction , $<$5\%, compared  to   the proportion  of   Sgr members  in   the
reference region, $\approx$25\%.  The individual  Sgr isopleth maps are then
constructed  by subtracting a suitably scaled,   and smoothed version of the
comparison  map  from the  reference    map.  The extra  smoothing  of   the
comparison map (only), at a scale of $\approx$30 arcmin, is used to suppress
additional  quantization  noise being added    in quadrature to the  desired
signal.

The Galactic gradient in these fields  is very large.  For example, counting
all stellar  images from  $B_{\rm J} =  18$ to  the plate  limits, typically
reveals an image density ranging from 10~arcmin$^{-2}$  at one plate edge to
20~arcmin$^{-2}$ at the opposite edge.  Of these on average around 10\% will
be Sgr members, mostly belonging to the main sequence.

The  effective resolution of the isopleth  maps at $\approx 10$~arcmin, is a
compromise between the desire to probe the detailed structure of Sgr and the
need to keep  the quantization noise  at  a manageable level.  However,  the
much larger  numbers  of Sgr  main  sequence members per unit  area (contour
intervals are $\approx 1$~arcmin$^{-2}$)  compared to the earlier red  clump
isopleth  maps, enable us  to probe much   further out along  both major and
minor  axis  directions, due  to the  reduction in random  noise.  There are
various systematic effects  present that preclude  combining all four of the
individual maps in a straightforward quantitative fashion, including lack of
deep  photometric calibrators  and  complex  systematic  field  effects  (in
addition to vignetting) present  on the plates   at the $\pm  0.1$~magnitude
level.  This coupled with the natural variation in plate limiting magnitude,
make selecting a consistent sample of Sgr main sequence stars very difficult
and help to explain why the individual isopleth maps outlined in Figure~1 do
not join  up  smoothly.  In spite  of  these caveats  the maps are  accurate
enough  to reveal the  general shape and extent of  Sgr, although we caution
that the full extent of  Sgr has to be greater  than the apparent boundaries
in the combined isopleth map.

Combining all this star  count information, \sgr\  has been detected down to
$b=-26^\circ$ along the E side of its major axis.  Fields further out do not
show any significant detection of stars which are likely to be main sequence
members  of \sgg.  The  last detected isopleth  in Figure~1 corresponds to a
surface density  of $\approx 1$ arcmin$^{-2}$,  which is a factor  $\sim 10$
below the peak surface density near  M54.  Isopleths constructed from either
main-sequence or  red horizontal branch stars are  consistent  with a system
which is to  first order symmetric about  a line of constant  latitude drawn
through the location of  M54 $(\ell=5.6^\circ,b=-14.0^\circ)$. The detection
of the DUO  RR~Lyrae stars in \sgg\ within  $\sim  5^\circ$ of the  Galactic
plane is consistent with this  symmetry, although stronger statements  await
better  calibration of the relative  surface  densities of RR~Lyrae and main
sequence stars in  the different fields studied.   Along the minor axis, our
isopleths  derived  from  main  sequence   stars  provide a   detection from
$l=3^\circ$ to $l=11^\circ$ degrees.  Thus, \sgr\  has been detected over an
area of   $22  \times 8$    degrees in  extent,   encompassing all  previous
measurements. This area includes the positions of  the four globular cluster
members.    We caution however,  that  to obtain  these  measurements of the
extent and shape of \sgg, we have had to patch together different data sets,
each    obtained with different  selection   criteria,  and most involving a
difference  between regions  of color-magnitude  space  which isolates \sgg\
only in   a statistical way.   The   shape,  extent, and   particularly  the
normalization of the contours that we have derived are therefore necessarily
uncertain.   Additionally,   it should    be  noted  that   the  statistical
significance of  the high frequency structure  in the isophotal  contours is
marginal.   Real substructure is  not physically  plausible, given the short
internal crossing times.

The apparent  flattening  of   \sgg\ is   $\sim$3:1,  corresponding  to   an
ellipticity of $\sim  0.7$.  Thus \sgg\ has significantly   higher projected
ellipticity than  the other  dSph  companions to the  Milky Way,  which have
projected ellipticities   $\simlt 0.3$ (with  the   exception of Ursa Minor,
which  has  ellipticity  $\sim  0.5$,  \markcite{irw95}Irwin  \& \des 1995).
Indeed, \sgg\ is as flat as the most flattened elliptical galaxies.

\subsection{Surface Brightness Profile of \sgg}

The  most robust, model-independent  quantities that  can be determined from
the isophotal  contours  are the radius along  the  minor axis at  which the
surface brightness has fallen to half of its central value, $R_{HB}$ and the
limiting radius, $R_{lim}$, again along  the minor axis.  The contours shown
here  give  values   of $R_{HB}   \sim     1.25^\circ$, and  $R_{lim}   \sim
4^\circ$. The half-brightness radius is a  lower limit to the characteristic
radius  of a  King  model fit,  $r_o$  (see, \eg  \markcite{bin87}Binney  \&
Tremaine 1987) and the limiting   radius will approximate the tidal  radius,
$r_t$.  Using the ratio $R_{HB}/R_{lim}$ to  estimate the corresponding King
model concentration parameter gives $c  \def \log{(r_t/r_o)} \sim 0.5$. This
is a typical value  derived for the dSph  companions to the Milky Way (Irwin
\&  \des 1995, their  Table~4 --- Carina, Draco,  Leo~II and  Ursa Minor all
have $c \sim 0.5$).

The total   luminosity of \sgr\  has  been estimated  by  \IGItwo\  from the
relative frequency of  the  RHB, giving $L_V   \sim  10^7 L_\odot$,  and  by
\markcite{mat96}Mateo  \etal   (1996), who derive   $L_V \sim  2 \times 10^7
L_\odot$.

\subsection{Distance and 3-dimensional Shape}

\subsubsection{Line-of-Sight Depth}

\paragraph{Field Stars}

The dispersion among the apparent  magnitudes of the RR~Lyrae stars detected
in  \sgg\ may    be used  to give  an   upper  limit   to the  line-of-sight
depth.   Alard  (1996)  finds    that  the extinction-corrected    magnitude
distribution    of  RR~Lyrae  variables may   be   fit  by  a Gaussian  with
$\sigma=0.3$ mag. This  dispersion  contains contributions from  photometric
errors, uncorrected  local variations in extinction and  a possible range in
RR Lyrae metallicities in the  sample.  Formally, these effects explain  the
whole observed dispersion.  However, interpreting the FWHM (2.34$\sigma$) of
this  distribution as an absolute  upper  limit to the differential distance
modulus of  \sgg\  yields an  upper limit to   the line  of sight extent  of
$\simlt 8 \kpc$ (for an assumed mean distance of 25~kpc).

Another estimate of the line-of-sight depth at a  point can be obtained from
the magnitude   spread of the   more numerous red  clump stars  over a small
field.  Stars   belonging  to the   red  clump  in   the  CMD  represent the
core-He-burning phase   of intermediate-mass and/or intermediate-metallicity
stars   (\eg \markcite{chi92}Chiosi, Bertelli   \&   Bressan 1992).   Some
intrinsic spread  in the  red  clump is expected, even    in a uniform  age,
uniform  abundance stellar population  (see \markcite{lee94}Lee, Demarque \&
Zinn 1994,  their   Figure~1). Nevertheless, even   in  a mixed  age,  mixed
abundance population the magnitude  spread across the  red clump provides an
upper   limit to  the distance   spread along  the line of    sight. We have
extracted  stars  of the appropriate color  and   magnitude to be  red clump
giants in  \sgr\ from the  CCD data of  \markcite{sar95}Sarajedini \& Layden
(1995); these data  are  from a  $7' \times  7'$  field, so problems  due to
patchy extinction and variations in  mean distance of  these stars should be
minimized.

Figure~2 shows the  distribution in magnitude of  field stars well away from
the line-of-sight to the globular cluster M54 (projected distance of $12'$),
and selected to lie in the narrow color strip $1.05 <  V-I < 1.17$. There is
a narrow, local maximum in the bin centered on $V =18.25$. This distribution
is modeled in the range $18.0 < V < 18.5$ as follows: the smooth and rising
distribution  of  `non-red-clump' stars (foreground  Galactic  stars and Sgr
giants and subgiants)  is fit by a  straight line segment  to a 0.2 mag wide
bin near $V=18.0$ and $V=18.5$, while the red  clump `hump' is modeled with
a Gaussian superimposed on  this sloping background.  The maximum-likelihood
fit  of the   (unbinned) data to  this  combined  model is superimposed   on
Figure~2; the Gaussian has a dispersion of $0.04$ magnitudes.
 
The width of the apparent magnitude distribution of these stars results from
a convolution of  intrinsic spread  due to   distance, age  and  metallicity
differences, together with the photometric error ($\sigma_V=0.01$ near ${\rm
V = 18}$), so it  is an upper limit to  the intrinsic dispersion in distance
modulus. Adopting a conservative  dispersion of 0.04~mag,  giving a FWHM  of
0.1mag, the corresponding dispersion in line of sight distance, again for an
assumed mean distance of 25~kpc, is $\Delta D=1.2$~kpc.

\paragraph{The globular clusters}

There   are 4 globular clusters  unambiguously  belonging  to \sgg, based on
commonality of  distance, radial  velocity   and celestial coordinates  (see
Table~1; references are given in the text where appropriate).  The variation
in the cluster distance  moduli can be used to  derive an upper limit to the
line-of-sight depth of  \sgg.  \markcite{dca95}Da Costa \& Armandroff (1995)
derive  the following  reddening-corrected distance  moduli: M~54 --  17.14;
Ter~8  --  16.69;  Ter~7   -- 16.72 and   Arp~2   -- 17.34.
\footnote{We  have corrected the distance modulus  for Arp~2 to take account
of the  fact  that the   apparent magnitude of  the  Horizontal  Branch that
\markcite{dca95}da  Costa  and    Armandroff  used  was   the   ZAHB  value,
V$_{HB}=18.30$, as given by \markcite{buo95a}Buonnano {\it et al.}  (1995a),
whereas, as pointed out by  \markcite{cha96}Chaboyer, Demarque \& Sarajedini
(1996), the mean value of the apparent magnitude of the horizontal branch is
the quantity that should be compared with the theoretical absolute magnitude
of the HB quoted (which  is not the ZAHB).  This  mean value is given by the
last authors as 18.18.}
These values   provide a mean  cluster  distance  modulus of   17.00, with a
dispersion of  0.32~mag, corresponding to $25\pm4$~kpc.   The  same range of
distance  moduli   (with   a   small   zero-point   offset)  is   derived by
\markcite{cha96}Chaboyer, Demarque \& Sarajedini (1996) for Ter~7, Ter~8 and
Arp~2  (they do   not have  M~54   in  their sample).  This  dispersion   is
remarkably similar to the spread in RR~Lyrae apparent distance moduli (Alard
1996), albeit that the globular clusters are spread over $\sim 8$~degrees on
the sky.

Thus,  the globular cluster distance data  are  consistent with those of the
field stars, but do not provide more precise constraints with extant data.

\paragraph{Three-Dimensional Shape}

The red clump  stars provide the  most robust estimate  of the line-of-sight
depth,  and from   above the   half-brightness depth  is  1.2~kpc.   This is
remarkably similar to  the minor axis parameters  derived  from the isopleth
maps of \S2.1, where the half-brightness minor axis diameter, for a distance
of 25~kpc, is $ 2 \times 550$~pc.  The 3-dimensional  shape of the Sgr dwarf
is thus  a   prolate  spheroid with  axis   ratios 3:1:1  and  a   long axis
approximately in the plane of the sky and at constant longitude.

\subsubsection{Mean Distance Across the Face of the Sagittarius Dwarf Spheroidal}

\IGI\ and \IGItwo\ derived the  distance to \sgg\ by  assuming that the  red
clump stars have the  same intrinsic luminosity   and color as those in  the
Small  Magellanic Cloud (SMC) and  matching SMC and Sgr color-magnitude data
to derive   the relative  distance  and reddening.  Adopting 57~kpc  as  the
(standard) distance for  the SMC provides a distance  to \sgg\ of $24 \pm  2
\kpc$, corresponding to a distance of the center of  \sgg\ from the Galactic
center of  $16 \pm 3 \kpc$ for  an adopted Solar  Galactocentric distance of
8~kpc.

A  similar estimate results from  analysis of the  apparent magnitude of the
$\sim  10$   RR~Lyrae  member  stars identified   by  the OGLE  microlensing
experiment group \markcite{mat95a}(Mateo \etal 1995a; \markcite{mat95b}Mateo
\etal 1995b). The significant improvement  in statistics provided by the DUO
microlensing   experiment, which has  identified  313 RR~Lyrae candidates in
\sgg\ (Alard 1996), again yields a mean distance of $24 \pm 2 \kpc$, for the
field  centered  at  $\ell=3^\circ$, $b=-7^\circ$,  for an  adopted RR~Lyrae
({\it ab} type) absolute magnitude $M_{B_J} = 0.79$.

In principle the most  precise distance at present can  be derived  from the
combination of the DUO RR~Lyrae data and  the distances to the four globular
cluster members (modulo the  line-of-sight depth for  the cluster system, as
discussed below). In the  case of the RR~Lyraes, however,  one is limited by
their unknown chemical abundances and, in both cases, one is further limited
by   lack  of knowledge  of   the  appropriate chemical abundance-luminosity
calibration for  horizontal   branch  stars,  since  the   internal chemical
abundance range in \sgg\ apparently spans at least 1.5dex. The most detailed
discussion is in \markcite{sar95}Sarajedini \& Layden (1995, their Table~3),
where they derive distances for M54 and for the field stars of \sgr\ under a
range of adopted relationships for  the absolute magnitude of the horizontal
branch. They obtain distances covering the range from 25~kpc to 29~kpc, with
the closer distance preferred  by the most  recent HB calibrations. The same
range of distances is derived from main sequence turn-off fits for Sgr stars
by \markcite{fahl96}Fahlman \etal (1996).

Presently,  there are too  few confirmed RR~Lyrae  members to provide useful
information  on  differential distances  over the dwarf.  However, red clump
stars are numerous  in \sgg, and  can be used  as standard candles, assuming
that there are no systematic variations  in stellar population with position
over that   galaxy. Twenty-six fields  along  the major axis  of  \sgg\ were
observed in V and I on  the 1.5m telescope at CTIO  (27 March 1996) with the
Tek1024 CCD. A  thorough description of the reductions  and analysis of these
data will be presented  in a subsequent  contribution. However, it is useful
for   the    present  discussion  to    mention    the  resulting   distance
constraints. Figure~3 shows the color-magnitude diagram  of one of the major
axis fields  at  ($\ell=15.48^\circ,b=-15.09^\circ$). Assuming that  most of
the reddening is due to foreground dust, the difference in reddening between
a given field and the template field (chosen  to be the off-cluster field of
\markcite{sar95}Sarajedini \& Layden 1995)  can be estimated by  finding the
difference in ${\rm V-I}$ color of the blue edge of  the disk sequence (seen
clearly in Figure~3 at ${\rm  V-I \sim 0.8}$, ${\rm  V < 17.5}$). All of the
photometry was extinction-corrected  in this  way to  the extinction of  the
template field. The V magnitude of the  peak of the red  clump was fit using
the technique described for Figure~2.  The uncertainties of these magnitudes
were   derived from bootstrap    resampling  and  include the    statistical
uncertainties  in   the extinction  correction.  Assuming  ${\rm  M_V(RR)  =
0.15[Fe/H] + 1.01}$ (\markcite{carn92}Carney, Storm  \& Jones 1992), and the
following parameters from \markcite{sar95}Sarajedini \& Layden (1995): ${\rm
[Fe/H]_{Sgr} = -0.5}$,  ${\rm M_V(red~clump) = M_V(RR) +  0.11}$ and  $A_V =
0.403$ (for  the template field) gives  the heliocentric distances displayed
in Figure~4.

A least squares fit of a straight line to the data presented in Figure~4 has
a slope  of  $0.03  \pm 0.05  {\rm  kpc/degree}$,  so an   adequate  working
statement for  present use  is that the  extant data  are consistent with  a
distance to \sgg\ of $\sim 25\pm2$~kpc, with no significant evidence for any
variation in distance along the major axis.

This value  of the distance  translates  the angular  half-brightness radius
estimated above  into a linear half-brightness  radius on the minor  axis of
$R_{HB} \sim 550$~pc.  Similarly, the limiting minor axis radius is $R_{lim}
\sim 1.7$~kpc. \Sgr\ is visible over a region with diameter $22^\circ$ along
the major  axis   and  $8^\circ$   diameter   along the  minor  axis.   At a
heliocentric distance of $25 \kpc$ this corresponds to a major axis diameter
of $9.6 \kpc$ and a minor axis diameter of $3.5 \kpc$.

\subsection{Metallicity and Age}

The discovery observations of \IGI\  allowed them to derive  (low-precision)
metallicities for Sgr K and  M giants from the equivalent  widths of the Mgb
feature. They determined the mean [m/H]  to be $\approx -1.1$. The existence
of an intrinsic metallicity  spread  was  deduced from  the  color-magnitude
diagram  by    \IGItwo, who  also deduced    that  \sgg\  has  a significant
intermediate-age stellar population based on the strong red clump and on the
existence of carbon stars.

\markcite{mat95a}Mateo  \etal\ (1995a)  confirmed, from the  distribution of
the  turn-off stars in their  deep (${\rm V-I}$,V)  data,  that the dominant
stellar population of Sgr has an age of 10~Gyr.  They deduced from the color
of  the  giant branch   that  the  mean metallicity    is ${\rm [Fe/H]  \sim
-1.2}$. \markcite{fahl96}Fahlman  \etal (1996) were  able to  fit their more
precise photometry   of main-sequence stars in   \sgg\ with isochrones which
range, in a  coupled determination of  abundance and age ([Fe/H];T), between
(${\rm [Fe/H] = -0.8};T = 10 \Gyr$) and (${\rm [Fe/H] = -1.2};T = 14 \Gyr$).

\markcite{sar95}Sarajedini \& Layden (1995) identified, in their photometric
study, three different  stellar populations in  the Sgr  dwarf. These  are a
metal-poor component,   typified  by M54, with  ${\rm   [Fe/H] =  -1.8}$; an
intermediate-metallicity  component  with ${\rm  [Fe/H]    = -1.3}$, and   a
dominant, relatively metal-rich component with ${\rm [Fe/H] \sim -0.5}$.

The Sgr globular clusters all have spectroscopic metallicity determinations,
(see Table~1) in each case  based on the  infrared calcium triplet strengths
for $\sim 5$ evolved stars. These metallicities  are: M54, [Fe/H$] = -1.55$;
Ter~8, [Fe/H$] = -1.99$; Ter~7, [Fe/H$] = -0.36$ and Arp~2, [Fe/H$] = -1.70$
(eg.   \markcite{dca95}da Costa \&    Armandroff   1995).   An   alternative
(photometric)  metallicity determination  for  Ter~7   is [Fe/H]$ \sim   -1$
\markcite{buo95b}(Buonanno \etal 1995b; \markcite{sar96}Sarajedini \& Layden
1996).   [This is  an  example of  the  fact that empirically, spectroscopic
metallicity determinations using  the Ca triplet have   been found to  agree
with  those based  on the  giant branch  color  only for metallicities below
one-tenth of the solar value; see Buonanno \etal.]  Adopting the photometric
metallicity for Ter~7  results   in a  mean  (unweighted)  globular  cluster
metallicity of $\sim  -1.6$~dex, some $\simgt  0.5$~dex more metal-poor than
the field star population, and a spread of $\sim  1$~dex, similar to that of
the  field   stars. (M~54     dominates a   luminosity-weighted  metallicity
distribution for the globular clusters, giving  a mean that is only slightly
more metal-rich than the unweighted mean.)

For comparison,  the  Fornax dwarf spheroidal galaxy   has a mean field-star
metallicity   of  [Fe/H$]  =  -1.4$  \markcite{buo85}(Buonanno  \etal\ 1985;
\markcite{beau95}Beauchamp \etal\ 1995), $\sim 0.4$~dex more metal-rich than
the average of  its cluster population,  [Fe/H$] = -1.8$. This difference is
approximately the same as that  discussed above for \sgg.  The full range of
the    field   star    metallicities   in      Fornax  is  $\sim     0.8$dex
(\markcite{beau95}Beauchamp \etal\ 1995), approximately the  same as that of
the five globular clusters  of Fornax (see  also \markcite{dca95}da Costa \&
Armandroff 1995), and again similar to the  metallicity spreads in the field
and globulars of Sgr.   The mean age  and age range  of the field  stars and
globulars   of Fornax  are   considerably  different  from  those   of \sgr,
however.  All studied globulars of  Fornax are consistent   with an old age,
while the Fornax  field stars show a dominant  population that  is less than
$\sim 5$~Gyr old,  with a range  extending  down to $\sim 1$~Gyr  (Beauchamp
\etal\ 1995).

The presence of RR~Lyrae stars in a stellar system is an indicator of an old
population. Theoretical quantification of `old',  by application of  stellar
evolution models   to horizontal  branch   morphology, remains    uncertain.
Observational calibration, by the detection of RR~Lyrae  stars in a range of
globular clusters  for   which  relative age estimates   are   available, is
feasible (\cf \markcite{ols87}Olszewski,  Schommer \& Aaronson (1987) for an
earlier  discussion of this approach).   IC~4499  is the youngest metal-poor
globular cluster ([Fe/H] $\sim -1.5$)  with a very substantial population of
RR~Lyrae variables;  indeed   IC~4499  has one   of  the   highest  specific
frequencies of RR~Lyrae  stars  \markcite{sun91}(Suntzeff,  Kinman  \& Kraft
1991;  \markcite{sar93}Sarajedini 1993).  On   the age scale  which has  the
canonical `old' halo being in the mean 17~Gyr old \markcite{cha96}(Chaboyer,
Demarque  \& Sarajedini  1996, their   favored calibration of  the RR  Lyrae
luminosity being $M_{V,\,RR} = 0.20{\rm [Fe/H]}  +0.98$), IC~4499 has an age
of     $\sim      12$~Gyr,  in  agreement     with      the   conclusions of
\markcite{fer95}Ferraro \etal\ (1995).  The Sgr clusters with ages estimated
by  \markcite{cha96}Chaboyer \etal\ are Ter~7,  $\sim 9$~Gyr on their scale;
Arp~2, $\sim 14$~Gyr; and Ter~8, $\sim 19$~Gyr.   Arp~2 has 2 RR Lyrae stars
detected, plus  12   HB stars  blueward of  the  instability strip  and none
redward      (\markcite{buo95a}Buonanno         {\it   et    al.}    1995a).
\markcite{ric96}Richer {\it et  al.} (1996) find that  Arp~2 is $\sim 5$~Gyr
younger than their fiducial globular  cluster at that metallicity  (NGC~7492
with [Fe/H]$ = -1.5$  and an estimated  absolute age of 15~Gyr), while Ter~7
is $\sim 4$~Gyr younger than its  fiducial at that metallicity (NGC~104 with
[Fe/H]$ = -0.7$ and an estimated absolute age of 14~Gyr).

A robust conclusion, consistent with both analyses of globular cluster ages,
is that    an  RR~Lyrae population  can  occur   in   stellar  systems  with
metallicities $\sim -1.5$~dex that are as much  as $\sim 5$~Gyr younger than
the classical old halo. Thus  the detection  of  RR~Lyrae stars in \sgr\  is
indicative of a population of age greater than $\sim 10-12$~Gyr.

In summary, the extant  data show that  globular clusters of \sgg\  are more
metal poor  than its field stars,  but cover a  very wide age range. The Sgr
field stars are predominately of   comparable age  to the younger   globular
clusters,  $\sim  12$Gyr, but  extend to sufficiently    young ages to allow
carbon stars to exist and to sufficiently old ages to allow a substantial RR
Lyrae  population.  The  field stars   also   cover a  very  wide  range  of
metallicity.   The distribution function   of metallicities  in  Sgr remains
poorly  determined. Most observations, however,   determine a mean abundance
more metal rich than $-1$~dex, with a range of  $\sim 1$~dex. \Sgr\ contains
the most metal-rich stars of any of the dSph companions of the Milky Way.

In \S5 we use these limits on  the age of the stellar  populations in Sgr to
constrain its survival against destruction by Galactic tides.

\subsection{Radial Velocities}

Two complementary sets of kinematic data for member stars in \sgg\ have been
obtained.

Spectroscopic data  from three observing  runs  on the 3.9m Anglo-Australian
Telescope (AAT)  determine the  mean kinematics of  \sgg.   The first two of
these runs (9--13 July 1990,  3--8 August 1993) were  part of a survey which
aimed to determine the kinematic and abundance distributions of the Galactic
Bulge.  The goal of the third AAT run (1--4 August 1994) was specifically to
measure  the gradient in  the mean line-of-sight velocity  over  the face of
\sgr\   and  to   provide   confirmed  member  stars  for  higher-resolution
spectroscopic study at  CTIO. The candidate  Sgr giant and asymptotic  giant
branch members were  selected from  CMDs  constructed from scans of   United
Kingdom Schmidt Telescope IIIaJ and IIIaF  survey plates using the Automatic
Plate Measuring (APM) machine \markcite{kib84}(Kibblewhite  \etal 1984).  In
sparsely populated  regions, such as  around Ter~8, all stars whose location
in the CMD was consistent with that  of the Sgr  giant branch were observed.
The   AAT observations used    the  AUTOFIB multi-object fiber  spectrograph
system.   A  detailed description of the   observational   details, the data
reduction and  resulting velocity and abundance   precision, together with a
detailed description  of     the    APM  photometry, is     presented     by
\markcite{me95a}Ibata \& Gilmore  (1995a). The velocity data resulting  from
the July  1990 and August 1993 runs  are listed in  \markcite{me95a}Ibata \&
Gilmore (1995a),  their Table~B1.  The data from   the August 1994   run are
presented  in Table~2.A, which  contain accurate coordinates, APM photometry
and  AAT heliocentric radial  velocities for 229 stars.  Repeat observations
of the  same star   are indicated  by  a  ``ditto''  in the   coordinate and
photometry columns.

Since the internal velocity dispersion of dSph galaxies is comparable to the
accuracy  provided  by AUTOFIB (the  velocity accuracy  of  the  sample from
\markcite{me95a}Ibata  \&  Gilmore 1995a  is  $\sim 12\kms$),  complementary
observations were obtained with the ARGUS multi-fiber setup  at the 4m Cerro
Tololo Interamerican  Observatory  (CTIO)   (2-6 August 1994)   feeding  the
bench-mounted  echelle spectrograph. The  advantage  of this instrument over
AUTOFIB is  a significant increase in the   precision of individual velocity
measurements,  which  however comes  at the   cost  of lower throughput  and
reduced      multiplexing    (24   fibres       instead    of    64).    See
\markcite{sun93}Suntzeff \etal  (1993)  for  a  comprehensive description of
ARGUS and  ARGUS data  reduction  procedures.  The Blue  Air Schmidt camera,
5186\AA\   filter  and   $1200\times200$ CCD   were  used   for  the echelle
observations.   Due to a  malfunction in the  CCD controller, the read noise
was exceptionally high, $6.8 \pm 0.6 {\rm e^-}$, resulting in an average S/N
$\sim 10$ in  the $\sim 1$ hour exposures.   Stars from M4,  M11, 47~Tuc and
NGC~288,  in  addition to the  IAU   velocity standard   star BS~9014,  were
observed for use as velocity templates.

The  resulting velocity  data  are  presented in   Table~2.B, which contains
accurate coordinates, APM photometry,   CTIO heliocentric radial  velocities
and Tonry-Davis crosscorrelation  R  values (\markcite{ton79}Tonry  \& Davis
1979) for  270 stars.  Repeat observations of  the same star using different
fiber setups are  indicated by a ``ditto''  in the coordinate and photometry
columns.  Some spectroscopic  integrations were divided into two  exposures;
where  applicable, an extra  entry in the table  lists the absolute value of
the velocity difference between measurements on  the two exposures.  Entries
marked with ellipses denote unavailable data.

The fields observed extend $\sim 10$  degrees along the  major axis of \sgg\
and   are  identified in  Figure~5.  Identification  of    members of Sgr is
straightforward from the kinematic data, as \sgg\ is offset in mean velocity
from Galactic bulge stars --- the primary contaminant --- by $\sim 150 \kms$
(\cf Fig  4 of \IGItwo).  Thus,  we select as members of  the  Sgr dwarf all
giant stars with heliocentric  radial  velocities satisfying $100\kms  \le v
\le 180\kms$.

Eighty-seven stars  were  observed at both  CTIO  and the AAT  and provide a
check on both the velocity zero points and the internal precision of the AAT
data.  The distribution of  differences between the repeat  measurements was
found to be Gaussian.  The mean difference in the  sense (AAT-CTIO) is $+1.4
\kms$,  while the dispersion is  $6.3 \kms$, which is  dominated  by the AAT
measuring error.

\subsubsection{Kinematic Results}

\Sgr\ covers  such  a large   angular extent   that  there is   a small  but
systematic variation across  its surface in  the projected component of  the
space  velocity of the  Sun. We thus  convert the heliocentric velocities of
Table~2 to values which would be seen by an  observer at the position of the
Sun, but in the reference frame  in which the Local Standard  of Rest at the
Sun is orbiting the Galactic center with a velocity of $220 \kms$. We define
this to be `Galactocentric' radial velocity, $v_{\rm GAL}$.

The peculiar velocity of the  Sun relative to the Local  Standard of Rest is
assumed     to  be   $v_{\rm pec,\odot}=16.5     \kms$    in  the  direction
${\alpha}_{1900}=18.0^\circ$, ${\delta}_{1900}=30^\circ$.   Defining $v_{\rm
HEL}$ to be the observed heliocentric radial velocity of  a star at Galactic
coordinates $(\ell,b)$,  $v_{\rm  pec,\odot,(\ell,b)}$  to be  the component
along the line of sight to the star of the  peculiar velocity of the Sun and
$v_{\rm rot,LSR}=220 \kms$  to be the circular  velocity  of the LSR at  the
Sun's location, then the `Galactocentric' radial velocity of that star is
\begin{equation}
v_{\rm GAL} = v_{\rm HEL} + v_{\rm pec,\odot,(\ell,b)} 
+ 220 \kms \sin {\ell} \cos b.
\end{equation}

The Galactocentric radial velocity  data, plotted against Galactic latitude,
are  shown  in Figure~6. The radial  velocities  of the 4  globular clusters
which are members of Sgr are also indicated  in Figure~6 and lie well within
the intrinsic dispersion of the kinematic data for Sgr field stars.

To quantify the variations in mean velocity and velocity dispersion in these
data, each of the eight fields observed at the AAT,  four of which were also
observed  at CTIO, was considered  separately.  A maximum-likelihood routine
was used to fit Gaussian functions to the data in  each of these fields; the
uncertainties on  the mean   and  dispersion were calculated    by bootstrap
resampling of the data.   The fitted parameters for  each field are given in
Table~3. The histograms of  the CTIO velocity data  in fields f1, f5, f6 and
f7 are  shown  in  the upper  row  of panels  in Figure~7  with   the fitted
functions  superimposed.  The lower two   rows of  panels   in Figure~7 show
histograms of the AAT velocity measurements. The data  show, along the major
axis, an increase in the mean velocity of $\sim  10 \kms$ over the $5^\circ$
from field  f1 to f4 and then  an approximately constant mean velocity until
the end of the data in field f8. Though the data do not span a wide range in
distance  along   the minor  axis, there   is no   evidence  for significant
variations in mean velocity in that direction (see below).

The Gaussian  parameters  fit  to   these  data  with  their   corresponding
uncertainties are displayed in Figure~8 as a  function of position along the
major axis of \sgr\ and are listed in Table~3. The mean velocities are shown
in  the left  hand  panel of  Figure~8. In each  subfield, the  mean and its
uncertainty for the  AAT  and CTIO data  are shown  separately. The velocity
dispersions derived  from the CTIO echelle data  are given in the right hand
panel. The mean  velocity  shows  a smooth  rise  of  $17 \pm  5 \kms$  from
$b=-26^\circ$ to   $b=-20^\circ$   ($\sim~3~\kms$/degree) and   then remains
constant, within the uncertainties, to $b=-12^\circ$ (see also Table~3).

The central  velocity   distribution (f7)  is  fit  by  a  Gaussian   with a
dispersion of $11.4 \pm  0.7 \kms$. This  value is  consistent with the  $12
\kms$    dispersion  obtained  using    just    the  4  globular    clusters
(\markcite{dca95}da Costa \& Armandroff, 1995).

The velocity dispersion of members of  \sgr\ at $b =  -25^\circ$ is equal to
the  central velocity dispersion $\sigma  =  11 \kms$.  Indeed, the velocity
dispersion is invariant over the face of  \sgr\ to within the uncertainties,
as quantified in Table~3. Further, within the precision allowed by the small
sample, the distribution of velocities is Gaussian in each of our fields.

\subsection{Minor Axis Rotation}

Assuming that the line $\ell=5^\circ$ corresponds to the major axis of \sgr,
we investigate the evidence for rotation around this  axis (\ie  minor axis
rotation).   Such streaming has been    observed in Ursa Minor  (Hargreaves,
Gilmore, Irwin \&  Carter 1994b), probably an  effect  of preferential tidal
stripping.  The kinematic data of Tables~2.A,~2.B, and  Table~B1 of Ibata \&
Gilmore (1995a), were divided  up into three groups  spanning the  ranges of
Galactic latitude $b > -17.5^\circ$, $-22.5^\circ < b  < -17.5^\circ$ and $b
< -22.5^\circ$. These velocity data are displayed as  a function of Galactic
longitude in Figure~9. The best  fitting straight lines to  these $\ell - v$
data are superimposed  on the diagrams. The slopes  of these lines are $-0.6
\pm  1.0 \kmsd$, $0.9  \pm  2.3 \kmsd$  and $-3.5 \pm   2.9 \kmsd$. Thus, to
within the  uncertainties, the fits are   all consistent with  no minor axis
rotation. This result is insensitive to the choice of center of \sgr.

\subsection{Proper Motion}

First epoch plates were taken in the early 1950's as part of the Palomar Sky
Survey. The E (red) survey plates  are well matched to  recent UKST R plates
and provide the proper-motion information covering  a 30+~year baseline. The
UKST IIIaJ sky survey plates were used  to provide information on colors, to
help discriminate between foreground stars and stars likely to be members of
\sgr.

The preliminary  proper motion of \sgr\ has  been estimated with  respect to
Galactic foreground  stars (a full  account  of this  work will be published
elsewhere). At  $l = 5^\circ$, $b =   -15^\circ$, three Galactic populations
need  to be considered:  the Galactic  bulge, at   an effective distance  of
$\sim10 \kpc$,  and   thin  and   thick   disk stars  with  maximum   volume
contributions at distances of $\sim 2 \kpc$ and $\sim 6 \kpc$, respectively.

On these  photographic  CMDs,  the red  clump/RHB is   the best-defined  and
populated part of \sgr\ and was used to pick out  the members. The RHB has a
color equivalent to  K stars (B$-$V$ = 1.0$)  and, in order to minimize  the
need for complex color--  and magnitude--dependent  astrometric corrections,
the  reference objects  were  selected over   a narrow grid  of  colors  and
magnitudes centered around  those of the Sgr RHB.  K dwarfs in the  Galactic
bulge have R  magnitudes $>$20 and Bulge  K giants have R  magnitudes $<$15,
implying that neither population is a significant contributor. Likewise, the
thick disk  K   stars  do  not   make a   significant  contribution  to  the
numbers.  The  dominant Galactic component in  and  around the RHB stars are
thin disk K dwarfs with expected R magnitudes of 17--18.

The expected reflex solar motion for both disk K stars and members of Sgr in
this direction is parallel to  the plane of the Galactic  disk and, thus, is
perpendicular to the direction of  elongation of Sgr.   This means that  any
apparent motion of member stars parallel to the direction  of the major axis
of \sgr\ is dominated by the orbital motion of Sgr.

The Galactocentric transverse orbital   motion of  \sgr\  is $2.1   \pm 0.7$
mas/yr, \ie  $250 \pm  90$ km/s, towards  the Galactic  Plane.  In  terms of
Galactic [U,V,W] velocities  (with a convention of  U positive in the radial
direction towards  the Galactic   Center, V in   the direction  of  Galactic
rotation and W northwards)  and assuming that the  unknown proper motion  of
\sgg\ parallel to the Galactic Plane is zero, the  total space motion of the
Sagittarius  dwarf  is then [232,  0:,   194]. The U  and   W components are
well-constrained, with errors  $\pm$60 km/s, while a secure  value for the V
component awaits  the use of  an extragalactic  reference frame.  This  is a
total velocity amplitude of $\sim 300$~km/s.

\section{The Orbit of Sgr}

The kinematic data presented above allow a new determination of the possible
orbits of \sgg.  The important physical  parameters of \sgg\ which constrain
its orbit are:
\nl
(i) the distance, which is well-determined in the central field near M54 and
reasonably constrained across its face;
\nl
(ii) the spatial  variation of the mean  radial velocity, which is extremely
important and now moderately well determined;
\nl
(iii) the internal dispersion profile, now moderately well determined;
\nl
(iv) the proper motion, which now has a preliminary measurement.

The mass of \sgg\ is not  important for constraining  the orbit, provided it
is  low enough that  dynamical friction  has  not substantially modified the
orbit over an  orbital period.  This limit  is $4 \times  10^9 \msun$ at the
current Galactocentric distance; we argue below that this is satisfied.

In principle, the best way to calculate the orbit of \sgr\ is to construct a
self-gravitating N-body model  and to follow its  evolution in the  Galactic
potential. The model would be evolved until the present time, at which point
the projected velocities of the particles and projected density of the model
would be compared to   observations.  Successive refinements to the  assumed
initial   position, initial velocity and  initial   mass distribution of the
model would allow iteration  towards a best-fitting solution.  However, this
approach is beyond the scope of the present paper.

The present velocity and  distance profiles across the  major axis of  \sgg\
probe the projected velocity and distance to the center of mass of the dwarf
as   it proceeds along its  orbit.   We explore an  assumption which greatly
simplifies  the orbit detemination: that stars  in \sgg\  can be regarded as
test particles which all move on the  same orbit. Numerical simulations (see
\S4 below) show that Galactic tides force a  satellite galaxy into a prolate
shape, and at pericenter  the satellite's longest axis  is aligned along its
orbit. However, this alignment with the orbit is only approximate, since the
internal  self-gravity in the  dwarf will act to  decelerate stars that lead
the  center of mass  to  lower energy orbits, and   to accelerate stars that
trail the  center of  mass to higher   energy orbits.  The prolate  dwarf is
thereby rotated  about its center of mass  such that the  leading edge drops
towards the Galactic center (see e.g., Oh \etal\ 1995).   However, this is a
small  effect.   The  self-gravity  also  affects  the major   axis velocity
profile.   In the limiting   case,  \sgg\ could be   considered as  a rigid,
tumbling bar, in which  the apparent velocity  at each point along the major
axis is the sum the projected  velocity of the  center of mass and the local
streaming  velocity.  We  consider  the validity  of  our approach in  \S5.3
below, where we show the likely amplitude of the failures of our simplifying
assumption is small.  The mean velocity  profiles of our two extreme  models
turn out to be similar, which suggests that the velocity gradient across the
dwarf is primarily determined by the potential of  the Milky Way rather than
by the rigidity of the dwarf.

Following  \markcite{joh95}Johnston {\it et  al.} (1995), we model the Milky
Way galaxy  by the sum  of three rigid  potentials,  with the disk component
described by a Miyamoto-Nagai model \markcite{miy75}(Miyamoto \& Nagai 1975)
\begin{equation}
\Psi_{disk} = - {G M_{disk} \over 
(R^2 + (a + \sqrt{z^2 + b^2})^2 )^{1/2}},
\end{equation}
the combined halo and bulge by a spherical Hernquist potential
\markcite{her90}(Hernquist 1990), 
\begin{equation}
\Psi_{sphere} = - {G M_{sphere} \over (r + c)},
\end{equation}
and the dark halo by a logarithmic potential
\begin{equation}
\Psi_{halo} = {v^2}_{halo} \log (r^2 + d^2).
\end{equation}
In these expressions, $R$ and $z$  are in cylindrical coordinates, while $r$
is the radial   distance in spherical   coordinates; $M_{disk} =  1.0 \times
10^{11} M_{\odot}$, $M_{sphere} = 3.4 \times 10^{10} M_{\odot}$, $v_{halo} =
128$~km/s, $a = 6.5$, $b = 0.26$, $c = 0.7$ and $d = 12.0$, all in kpc.

We will first obtain constraints  on the orbit of  \sgg\ assuming that there
are no systematic    internal  streaming motions    (essentially  neglecting
rotation and  expansion) or   figure  rotation (tumbling).  With this,   the
apparent mean  velocity at each  point along the   major axis of  Sgr may be
modeled by the true space velocity of a test  particle at the corresponding
point on the orbit.

The test particle is started off at  the present position  of M54, using $25
\kpc$  as the value for the  distance of \sgg\  at this point. We explicitly
assume zero transverse velocity perpendicular to the direction of elongation
of Sgr, as  in \S2.7 above.   The other two  components of space motion  are
specified, by initial guesses. The equations  of motion are integrated along
the   orbit using  a     Runge-Kutta scheme.   The   `amoeba'  routine    of
\markcite{pre86}Press  \etal (1986) is  used to refine successive guesses of
the initial velocities, subject to the  constraint that the component of the
velocity of the test particle is consistent with the mean radial velocity in
the observed fields presented above.

Since the data are  binned, it is natural to  use the $\chi^2$ statistic for
this comparison:
\begin{equation}
\chi^2 = \sum_{fields} (v_i-v_{model})^2/{\delta v_i}^2,
\end{equation}
where  $v_i$ is the    mean velocity in field    $i$, ${\delta v_i}$  is the
standard  error on $v_i$  and $v_{model}$ is the   projected velocity of the
test particle at the point in its orbit that corresponds  to the position on
the sky of field $i$.

The orbit of the test particle which minimizes $\chi^2$ as defined above, is
shown in  Figures~10  and 11.  The left-hand  panel  in Figure~10 shows  the
radial velocity data in the $b$ -- $v$ plane together with the test-particle
orbit. This fit is formally acceptable (reduced $\chi^2 = 1.7$), in spite of
the systematic differences between the model and the mean velocity data. The
curvature of the data near $b=-15^\circ$ is so extreme that  no orbit in the
above potential  can  be made to pass  through  the data points  better than
this.  The right-hand panel compares  the orbit to the heliocentric distance
measurements presented in Figure~4.  This fit is acceptable (reduced $\chi^2
= 1.8$).   Note that  the distances to  the RR~Lyrae  stars at  low latitude
(Alard 1996) discussed above in \S2.3.2 are not included;  we return to this
point in \S5.3 below.  Figure~11 shows the $x$  -- $z$ shape of the best-fit
center-of-mass  orbit,  integrated for $10^9$   years.  The radial period of
this     orbit  --- defined  as   the    time  taken from  apogalacticon  to
perigalacticon and back --- is $0.76$ Gyr.

We note  that  this orbit requires  that the  end of  Sgr farthest from  the
Galactic plane is  closest to the observer, by  an amount of $\sim  2.5$~kpc
per ten degrees along the major axis.  This prediction is testable by direct
determination of the distance to each end of \sgg.  Perhaps the most precise
such  determination, given the large  numbers  of identifiable member stars,
will be comparison of the mean apparent magnitude of the RR~Lyraes in Sgr at
either end of the major axis. A  large sample is  already available from the
DUO RR~Lyraes,   which are  near the   Galactic  Plane and, thus,   the more
difficult to detect.  Note further that this best-fit orbit  gives rise to a
proper  motion  of  $2.0   {\rm mas/yr}$,  towards   the  Galactic plane, in
excellent agreement with the measured value of $2.1 \pm 0.7 {\rm mas/yr}$ in
this direction.

\subsection{Major Axis Rotation?}

The mean  velocity at each point  along the major  axis of  \sgg\ is made up
from the sum of the velocity of a test particle at that  place moving in the
Galactic potential on the same orbit as the center of mass of \sgg, together
with  any contribution from systematic motions  associated with the internal
dynamics of \sgg. There are two  possible such systematic motions associated
with \sgg\ itself --- internal streaming and bulk figure tumbling.

The  numerical calculations summarized in \S4  below show that the effect of
Galactic   tides on a  dwarf spheroidal  is  to  generate flattening, and to
orient the dwarf  spheroidal such that its major  axis is aligned along  its
orbit, at  least  near perigalacticon  for elliptical   orbits.  That  is, a
tidally-distorted dSph  galaxy is  effectively   a bar which  is  spin-orbit
locked near perigalacticon.  In this  case, no bulk figure tumbling bringing
the major axis out of the plane of the orbit is expected.  In \S2.6 above we
placed tight limits on any minor axis rotation. Thus, we restrict discussion
here  to consideration of any  possible internal  streaming motions in \sgr\
involving rotation parallel to the apparent major axis.

Is it plausible that \sgr\  is a rotating body?  \markcite{har94a}Hargreaves
{\it et al.}  (1994a) found no significant rotation in the Sextans dSph, but
discovered  rotation  about  the     long  (major)   axis  of  Ursa    Minor
(\markcite{har94b}Hargreaves {\it et al.}  1994b).  They speculated that the
rotation in the latter   case was attributable to  tidally-induced streaming
motions.  In any case,  the  amplitude in  UMi is  a  small fraction of  the
internal   velocity dispersion.  Thus, based on   the precedent of the other
dSphs,  no significant internal streaming motion  is expected in \sgg. It is
probable  that the observed mean  velocity   is correctly representing  that
which   is  appropriate    to   determine  the  orbit   of   Sgr   about the
Galaxy.  Nonetheless,  we consider as  a  limiting case the possibility that
\sgg\  is flattened entirely by  internal  streaming, and that streaming  is
oriented so  as  to have  the maximum  possible effect on  the observed mean
stellar velocities.

The  Sagittarius dwarf is  significantly flattened, with 3-dimensional shape
$\sim$3:1:1.    \markcite{bin78}Binney     (1978) studied      models     of
self-gravitating  (with mass following light)  prolate galaxies flattened by
internal  rotation. We assume (maximally)  that the  apparent ellipticity of
Sgr  is the true  ellipticity, and that  the shape is determined entirely by
rotation,      with no Galactic       tidal   contribution. Then   Fig~2  in
\markcite{bin78}Binney (1978) indicates that, if rotationally flattened with
ellipticity $\epsilon = 1 - {a_1/a_3} \sim 0.67$, that the maximal streaming
velocity in  \sgg\ is approximately equal to   its velocity dispersion.  For
consistency  with the   model developed  below, we  further  assume that Sgr
rotates with  a solid body  rotation curve about  a point near M54.  In that
case,  the maximum rotation  velocity corresponds to  $|\Omega_{rot}| \lta 1
\kmsd$. We note that   a  robust test of   this model  will be  possible  by
determination of the mean velocity for the  DUO RR~Lyrae stars, which are at
the other end of the major axis of \sgg\ than the location  of the stars for
which we have extant kinematic data.

The orbital  calculations described in  the previous section were altered to
account  for such bulk rotation. The  radial periods resulting  from fits to
the data were lengthened to $1.0 \Gyr$ for $\Omega_{rot}  = 1 \kmsd$ (in the
sense that lower end of \sgr\ near Ter~8 is rotating away from the observer)
and shortened to $0.6 \Gyr$ for $\Omega_{rot}  = -1 \kmsd$. Thus, the period
of the orbit of \sgr\ cannot be greater  than $\sim 1.0  \Gyr$ even if it is
conspiring to rotate in such a way so as to counteract the observed velocity
gradient.   None of  the models  with major  axis rotation in  the range $-1
\kmsd <  \Omega_{rot} < 1 \kmsd$  fits the curvature  in the  velocity curve
near $b=-15^\circ$ evident in Fig~10.

Thus, the orbital period of  \sgg\ about the Galaxy  has a formal best value
of 0.76Gyr.    This orbit is  formally   an  acceptable  fit to  the  extant
kinematics, in   spite of  the apparent   systematic deviations.  No  orbit,
subject to the assumptions specified above, with period significantly longer
than $\sim 1$Gyr provides an acceptable fit to the data.   We note here that
it is implausible  that \sgg\ has  been recently captured  by the Milky Way.
For this to have happened, a non-destructive  encounter of Sgr with a rather
massive companion galaxy is required to perturb  its orbit sufficiently. The
apocenter we have derived is  similar to the  current distance of the  Large
Magellanic   Cloud, but  the current constraints   on  the orbit  of the LMC
(Jones, Klemola \& Lin 1994) are not consistent with such a scenario.

\section{Results of numerical simulations of the evolution of dwarf spheroidal galaxies}

All of the orbits calculated above have periods $\lta 1 \Gyr$, implying more
than ten  perigalacticon passages  over the last   $10 \Gyr$,  which  is the
minimum age  of  the dominant  population in   Sgr. \markcite{joh95}Johnston
\etal\  (1995)  and \markcite{vel95}Velasquez  \&  White (1995) attempted to
model the   orbital  evolution    of  \sgg\  using    the  \IGI\  data    as
constraints. Analyses of the tidal  effects on generic satellite galaxies by
\markcite{pia95}Piatek \& Pryor  (1995)  and  by \markcite{oh95}Oh,  Lin  \&
Aarseth  (1995)  are  also relevant   in considering   the survival  of  the
Sagittarius dwarf over its   lifetime, at least in  terms  of what  did {\it
not\/} occur.

The  conclusions in  common  in all  of these   studies  are that  the dwarf
galaxies in which mass follows light lose a very  substantial amount of mass
at every close perigalacticon passage.  The  internal velocity dispersion in
the satellite  does not    change  significantly prior to  disruption   (the
dispersion may in  fact decrease during disruption)  and the stellar surface
density distribution remains well described by a King model until near total
disruption. Thus,  there are no clear observational  signatures of the tidal
disruption of a  dSph  galaxy, prior to  its  final demise, except  the very
obvious one of  the  presence of  a substantial tail  of stellar  (and  dark
matter)  debris spreading along a  dispersion  orbit  with the remnant  dSph
galaxy.

\markcite{pia95}Piatek \&  Pryor (1995) modeled the  evolution of  a stellar
satellite galaxy with parameters similar to those inferred from the luminous
component of the  Ursa Minor dwarf spheroidal  --- masses in the  range $3-8
\times  10^5\msun$, no dark  matter   and  with  the stars  described  by  a
single-component King model,  with  core radius  (defined by them  to be the
radius at which  the surface brightness has  fallen  by a factor of  two) of
140~pc and  tidal   radius  of 643~pc.    The   orbits investigated   in the
simulations  had   perigalacticon   at 30~kpc,  and   various  apogalacticon
distances, but were always highly elliptical. A range of Galactic potentials
was assumed, from    the  extreme of  a   point  mass to an   infinite  dark
halo. \markcite{pia95}Piatek \&  Pryor  found that, generically,  the  dwarf
spheroidal is stretched along its  orbit and compressed in the perpendicular
directions.   The amplitude of  the  tidal  effects   is very  sensitive  to
perigalactic distance,  but only moderately  sensitive to the intrinsic mass
of the dwarf.  In  general, the apparent M/L   of the dwarf  changed  only a
little before complete disruption of the dwarf spheroidal, which took a very
small number, of order one, of perigalactic passages.

\markcite{oh95}Oh, Lin  \&  Aarseth (1995)  consider an extensive  series of
calculations  investigating the   evolution of   fragile systems  in   tidal
fields. They  also assumed as initial conditions  for the satellite galaxy a
single-component, isotropic King profile, with  core radius of a few hundred
parsecs  and,  in this case,  a mass  in  the range $2-6  \times 10^6\msun$.
Thus, again they simulated the evolution of a low-density stellar satellite,
with no dark  matter.   In agreement  with  \markcite{pia95}Piatek \&  Pryor
(1995),  they   found that   the  dwarfs   were  distorted  into a  triaxial
configuration, such that the longest axis is in  the direction of the orbit.
After a significant fraction of the original mass is lost, the flattening in
the orbital plane is preserved, but the major axis  becomes skewed, with the
leading end closer to  the Galaxy and the trailing  tail further away. After
disruption,  the stars which escaped  attain Galactic orbits similar to that
of the original parent satellite galaxy. When  a dwarf spheroidal is tidally
disrupted, the velocity  dispersion  of the unbound,   but not-yet-dispersed
stars is comparable to that internal to the dSph, in virial equilibrium just
prior to the disruption. Thus, an observed large velocity dispersion implies
the  presence of a large mass,  which  is dark  matter in the  case of dSph,
irrespective of whether or  not the stars are  in virial equilibrium  or the
dSph is being tidally disrupted.

\markcite{oh95}Oh, Lin \& Aarseth (1995) show that the  rate at which a dSph
galaxy is disrupted  is   a very  sensitive  function of  the ratio  of  its
physical size  to its   tidal   radius (see also   \markcite{joh96}Johnston,
Hernquist   \& Bolte (1996)   who characterized  survivability  in  terms of
density  contrast).  Dwarf spheroidal  galaxies whose initial extent is less
than  approximately twice their tidal  radius are typically  able to survive
for a Hubble time.

The above two sets  of simulations lead to  models which include dark matter
in   the       satellite  galaxy,   to    increase      its   survivability.
\markcite{vel95}Velasquez \&  White (1995) modeled  the orbital evolution of
the  Sgr dwarf, adopting  a King model for  the  dwarf with  a mass of $\sim
10^8\msun$,  a core radius of 527~pc  and tidal radius  of $2735 \pc$.  They
derived  a  period of order  0.75  Gyr and  perigalactic  distance  of $\sim
10$~kpc.  They found that a single passage though this pericenter completely
disrupts the dwarf,  unbinding  some 95\% of  its  mass within two   orbital
passes. They emphasize, however, that their orbit requires the Sgr galaxy to
survive at  least  10  pericentric passages  before   being  almost entirely
disrupted  on  the   passage  immediately  preceding the   present  one.  No
mechanism to ensure the survival the Sgr dwarf prior to its current orbit is
discussed.

\markcite{joh95}Johnston,   Spergel  \& Hernquist (1995)   also  modeled the
specific case of \sgg, this time adopting as a model a Plummer sphere (which
has a constant density core and $\rho \propto r^{-5}$  envelope) with a mass
of either $10^7\msun$ or $10^8\msun$  and a  scale radius  of 600~pc.   They
analyzed a variety of orbits with periods in the range 1.5--2.5~Gyr, none of
which is, however, consistent with the  observed low gradient in mean radial
velocity along  the  Sgr dwarf.   \markcite{joh95}Johnston \etal\ note  that
fitting the shallow  velocity gradient   discussed above requires    shorter
orbital periods and, with their adopted  model of \sgr, an unacceptably high
disruption   rate  of the  dwarf,  in  agreement   with the  conclusions  of
\markcite{vel95}Velasquez  \& White (1995).  \markcite{joh95}Johnston \etal\
(1995) considered in  detail the stability of  the  dwarf, as a  function of
orbital parameters. Even with their least disruptive orbits and more massive
satellite galaxy,  some 20\%   of  the original  satellite mass  is  tidally
stripped at each pericentric passage.  They confirm that  the effect of this
mass loss  on the current internal kinematics  of the dSph is  minimal, with
the   central velocity dispersion  and   its radial variation,  being almost
entirely unaffected. Similarly, no perceptable  deviations from the original
Gaussian velocity profiles are generated.

\section{Mass Density and Survivability of \sgr}

The simulations  discussed above obviously did  not model the case  of \sgg,
since it has persisted for many more orbital periods  than any of the dwarfs
in those  analyses.  Their  relevance for  the  case of  \sgg\ lies  in  the
understanding of  the prolate shape  of \sgr\  as  reflecting the effects of
Galactic tides.  We  propose below  that  \sgr\ is being tidally   stretched
along its major  axis, which  is aligned with   its orbit, and is  presently
tidally limited.  We present a self-consistent model with these properties.

\subsection{The Mass Profile of  \sgr}

The extant  simulations of the orbital evolution  of \sgr\ which allowed the
presence  of  dark  matter   --- namely  those  by \markcite{joh95}Johnston,
Spergel \&   Hernquist  (1995;   their  more  massive  satellites)   and  by
\markcite{vel95}Velazquez \& White (1995) --- were limited to ones where the
light traced the  mass.  These  simulations  concluded that  \sgr\ should be
rather easily disrupted, leading to the conundrum of its  age being an order
of magnitude greater than its  predicted lifetime.   How can one  circumvent
these contradictory  conclusions?   We begin  by considering the  properties
expected of a dark halo in a dwarf galaxy.

Gas-rich dwarf irregular galaxies have   HI rotation curves that are  rising
over much of the extent of the data (\eg \markcite{cari90}Carignan, Beaulieu
\&  Freeman 1990; \markcite{lak90b}Lake, Schommer \&  van  Gorkom 1990), \ie
solid-body rotation curves.  This suggests that they  are embedded in a dark
halo of almost  constant density, so  that the optical   galaxy --- and  the
more-extended HI gas disk  --- resides within  a dark halo whose core radius
is significantly larger than the extent of the luminous  galaxy. Fits to the
rotation curves of gas-rich dwarf galaxies which do  have a turn-over in the
rotation curve provide values of  the dark halo  core radius of around 3~kpc
(\eg       \markcite{lak90b}Lake,  Schommer     \&     van    Gorkom   1990;
\markcite{moo94}Moore  1994).  Dwarf  irregulars  are   significantly   more
dark-matter-dominated than are large spirals (\cf \markcite{cari90}Carignan,
Beaulieu \& Freeman  1990);   indeed lower  luminosity spirals may   be more
dark-matter    dominated    than    are     higher      luminosity   spirals
\markcite{per96}(Persic, Salucci \& Stel 1996).

Model-dependent determinations of the density of dark  matter in the core of
the gas-rich dIrr  galaxies  discussed  above  provide rather high   values,
$\simlt 0.1\msun$/pc$^2$. As emphasized by \markcite{moo94}Moore (1994), the
large core radii for the dark halos  around gas-rich dwarf galaxies are not
consistent with the  halos consisting of  Cold Dark Matter.   CDM-dominated
N-body  simulations predict rotation  curves which are significantly steeper
in the inner regions than those observed, even before  taking account of the
likely added CDM contraction after baryons settle  in the central regions of
the dark halo  (\cf \markcite{flo93}Flores {\it et  al.} 1993). Note that to
be  trapped at all  in such small-scale systems, the  dark  matter has to be
either very cold non-baryonic matter, or dissipative (hence baryonic).

We  propose that  the gas-poor dwarfs,  such  as the dSph,  have  dark halos
similar to  those of the  gas-rich dwarfs. This  is of course what one would
expect if  it   is  only star   formation  histories  and   gas   flows that
differentiate the two types  of galaxy.  The close proximity  of the dSph to
the Milky Way raises the possibility of tidal truncation of the halos of the
dSph,  leaving the  constant density  core (provided it   is of high  enough
density). Thus, we  consider the situation if \sgg\  stars are embedded in a
constant-density dark halo. We  will retain the   usual assumption that  the
system  is isothermal  ---  as   discussed above, the   observations of  the
velocity dispersion are  quite consistent with  no variation across the face
of Sgr --- and has an isotropic velocity dispersion  tensor.  Models of this
type  were     analyzed    by   \markcite{lak90a}Lake   (1990)     and    by
\markcite{pry90}Pryor  \& Kormendy (1990) in analyses  of the Draco and Ursa
Minor dSphs.

\subsection{The Survival  of \sgr}

An   isothermal   tracer population  embedded    within  a constant  density
(spherical)  dark halo (corresponding  potential $\Phi (r)  = 2 \pi G \rho_o
r^2$) will have phase space density
\begin{equation} 
f \propto \exp(-E/\sigma^2),
\end{equation} 
where $E  = \Phi  + {1 \over   2} v^2$ and  $\sigma$  is the one-dimensional
velocity  dispersion. The integral  of this expression over velocities gives
the volume density of the tracer stars,
\begin{equation}
\rho_{\ast}(r) = \rho_{\ast}(0)\exp(-\Phi/\sigma^2).
\end{equation}
Substitution   of  the expression above    for  the potential  (quadratic in
coordinate $r$), defining the characteristic scale $a$ such that
$a^2 = \sigma^2 / 2 \pi G \rho_o$, 
and integration down   the   line-of-sight yields the   observable   surface
brightness profile,   assuming no radial   gradient  in the  conversion from
stellar number  density  to surface   brightness, that  is, no radial   mass
segregation:
\begin{equation} 
I(R) = I_0 \exp (-R^2/a^2).
\end{equation} 
Thus the surface  brightness profile of an  isothermal  tracer population of
given velocity dispersion in a simple  harmonic potential (constant density)
is a Gaussian in the   projected distance from  the   center, with a   width
determined by the value of the constant density. This Gaussian has fallen to
half its central  value at $R_{HB}  =  \sqrt{\ln{2}} \,  a = 0.83\,  a$ and,
using the definition of $a$ above,
\begin{equation} (
G \rho_o)^{-1/2} = \sqrt{{4 \pi \over 4.16}} { R_{HB} \over \sigma}.
\end{equation}
As discussed by \markcite{lak90a}Lake (1990) and by \markcite{pry90}Pryor \&
Kormendy (1990), relaxing in  this manner the  requirement that light traces
mass (explicitly required in fits of one-component King  models to the light
profile) provides  a determination of  the central mass  density  which is a
factor of $\simgt 2$ {\it lower\/} than the King model value.

Note that, in principle,  this model may be tested  by this prediction  of a
Gaussian surface brightness  profile; the current data  for \sgr\ are not of
high  enough   S/N for     this  to  be      a constraint.  As    shown   by
\markcite{lak90a}Lake   (1990)  and   by  \markcite{pry90}Pryor \&  Kormendy
(1990),  the   fits  to Ursa  Minor  are   acceptable,  while some  velocity
anisotropy seems to be  required for Draco ---  the Gaussian fall-off is too
steep.  Recall, however, that an  initially isothermal velocity distribution
function which is truncated  by tides is  exactly a King  model, so that one
does not expect a pure  Gaussian model to be  a good  fit  to the data,  but
neither is adequacy  of fit  by a  King model an  argument  in favor of  the
assumption that mass traces light.

The minor-axis radius at which the surface brightness of \sgr\ has fallen by
a factor  of two  was derived above   to be $R_{HB}  \sim  550$~pc  (but, as
discussed, is uncertain due to problems with subtracting Galactic foreground
stars over a large range in Galactic latitude),  in very good agreement with
the independent estimate of the line-of-sight half-width.  The line-of-sight
stellar profile is the relevant quantity  for substitution into equation (9)
above     (\cf Lake  1990).   The    velocity   dispersion   of \sgr\     is
well-determined.  Thus a reasonable estimate  of the central mass density of
\sgr\ is
\begin{equation} 
\rho_o = 0.03~(550\,{\rm pc}/R_{HB})^2 (\sigma/11.4\, {\rm km
s}^{-1})^2 \msun ~ {\rm pc}^{-3}.
\end{equation}

The total  mass of the Galaxy  interior to the  present position of \sgr, $d
\sim 15$~kpc, for a flat rotation curve with $V_c=220 \kms$ is
\begin{equation}
M(<15{\rm kpc}) \sim 2 \times 10^{11} \msun,
\end{equation} 
giving a mean enclosed density of
\begin{equation}
<\rho> \sim 0.013  \msun {\rm pc}^{-3}.
\end{equation}
Note that the halo density evaluated {\it at\/} 15~kpc is 
\begin{equation}
\rho_{halo}(15{\rm kpc}) \sim 0.004  \msun {\rm
pc}^{-3}.
\end{equation}

Thus the density inferred for \sgr\ on the basis  of a constant-density dark
halo model  is  $\sim 3$ times  greater   than the mean  background  density
interior to pericenter.

The number of orbits that a satellite can remain intact against a background
tidal field increases  with the density  contrast between the  satellite and
the mean background density interior to pericenter of its orbit. Indeed, the
classical Jacobi/Roche criterion ---   which assumes point masses,  circular
orbits and phase-locking  of the  satellite in its  orbit ---  gives that  a
factor of three in mean density contrast  between the satellite and the main
galaxy is required for stability of the satellite.   Note that the classical
tidal  radius derived using   the  Jacobi  integral   is a  remarkably  good
approximation in many cases,  but recall the  result from Oh, Lin \& Aarseth
that satellites on  elliptical orbits could  persist provided their limiting
radius was less than  {\it  twice\/} their  tidal radius at  perigalacticon.
Satellites on elongated orbits are  expected to be  more robust, for a given
pericenter    (\cf  Allen  \&   Richstone  1988).  \markcite{joh96}Johnston,
Hernquist \& Bolte (1996) recently  investigated the evolution of satellites
in  strong tides and characterized  their longevity by  the ratio of central
density of the satellite to the mean background.   However, the mean density
of the satellite is the more meaningful quantity for such a comparison.  The
Plummer model profiles   adopted by \markcite{joh96}Johnston, Hernquist  and
Bolte have  a mean density  within  two  scale-lengths  (close to the  outer
detectability  limit) a factor  of $\sim 8$  below  the central density and,
hence, a factor of 8 below the mean density of a  uniform system, as modeled
here.  They found that a central density of  the satellite which is a factor
of  30  higher  than  the mean   background  interior  to perigalacticon was
sufficient for a satellite   to  survive.  This  translates into  a  density
contrast of $\sim 3$ as being the requirement  for a tidally robust, uniform
density satellite, as discussed here, in remarkable agreement with the Roche
criterion.

These results imply   that the structure we    propose for \sgg\ is   fairly
robust, and  also  that tides  have peeled  away   the outer, lower  density
envelope to the halo (otherwise the factor of three  is a mere coincidence).
\Sgr\ is not impervious to Galactic tides, and is  being tidally limited and
distorted, though not disrupted.

With the model of a constant-density dark halo, estimation of the total mass
of \sgr\  is straightforward, providing one can  argue  for a cutoff radius.
The tidal limitation suggests that one be guided  by the extent of the stars
down the minor axis.  Self-consistency of our picture also requires that the
mass be low enough that dynamical friction be ignorable.

The fits to gas-rich dwarfs discussed above found a core radius for the dark
halos of $\sim 3$~kpc.   The minor axis extent of  the stellar  component of
\sgr\ is $\sim 1.7$~kpc, and is a lower limit  to the possible extent of the
dark  halo in this  direction.  We shall adopt  2~kpc as a  fiducial for the
cut-off radius of the constant-density dark halo, $R_{Sgr,halo}$; one should
be aware of the dependencies of derived quantities on this value (we express
all quantities with this dependence explicitly).
\begin{equation}
M_{Sgr,halo} \sim { 4 \pi \over 3} R_{Sgr,halo}^3 \,  \rho_o,
\end{equation}
\begin{equation}
M_{Sgr,halo} \sim  10^9 \msun (R_{Sgr,halo}/2{\rm kpc})^3
(\rho_o/0.03 \msun {\rm pc}^{-3}).
\end{equation}

The  mean  orbital Galactocentric  distance derived  above is $\sim 30$~kpc,
thus
\begin{equation}
t_{dyn\, fric} \sim 1.3 \times 10^{10}\, {\rm yr} ({d \over 30\,{\rm
kpc}})^2 ({ 10^9 \msun \over M_{Sgr}}).
\end{equation}

Detailed numerical simulations are required   to test the conclusions  about
tidal  stability  and   orbital    evolution   based on   these     analytic
approximations.

\subsection{Consistency of the Model with the Derived Orbit}

In \S3 we derived the orbit of \sgg\ under  the assumption that the internal
binding force was  negligible. This assumption is, of  course,  at odds with
the  model  presented above,  which proposes the   existence of a dense dark
matter halo around the dwarf  to ensure its  survival against Galactic tides
until  the present  time.  Below,  we  therefore investigate  the  projected
distance and  projected  velocity of the stellar  component  of \sgg\ in the
presence of the proposed dark matter halo, to  check whether the assumptions
used in \S3 to calculate the orbit of the center of mass of \sgg\ still hold
to a reasonable approximation.

In the model presented  in \S5.2 above, the internal  dynamics of  \sgg\ are
completely dominated by a   dark-matter   halo, which, for  simplicity,   is
assumed to be spherical.  As the nature of dark  matter remains unknown, the
simplest assumption is that the dark  matter mass distribution does not vary
significantly  with time.  This means that  the  self-gravity of the stellar
component  is negligible, so  the orbits of tracer stars  in the dark matter
halo can be integrated without the need to resort to an N-body scheme.

A  spherical   isothermal population   of 10000   particles was  constructed
according to Eqns~6 and 7, and was placed in a dark halo of constant-density
$\rho_0=0.03 \msun  ~ {\rm pc}^{-3}$ and radius  1.4~kpc. The center of mass
of the dark matter halo  will follow some orbit  in the Milky Way potential;
for concreteness, suppose it  moves along the  orbit calculated in \S3. At a
given time $t$, a tracer-particle of the stellar population will find itself
at  a distance  $d$ from the  center  of mass of the  Sgr  dark halo. So  in
addition to   the acceleration due  to  the Galactic mass  distribution (the
assumed potentials are      given in \S3),  the  particles     experience an
acceleration
$a_{Sgr}=G {4\over 3}\pi \rho_0 d $ if $d < 1400\pc$ or
$a_{Sgr}=G M_{Sgr}/d^2$ if $d > 1400\pc$.
The  numerical  method used  to integrate the  orbits  was identical to that
described in \S3.

The integration was started  10~Gyr ago, and evolved  up to the present  ---
fully 12 orbits --- at which time the radial  velocity gradient and distance
were projected onto the line  of sight from the  Sun. Figure~12 compares the
radial   velocity  gradient and  the heliocentric   distances of the stellar
tracer particles in this model, with those based on the  orbit of the center
of mass  orbit.  Both techniques give predictions   that are consistent with
the observations, and in particular inclusion of the internal forces produce
stellar distances across the  face of \sgg\ that  may be in better agreement
with the RR~Lyrae distances at low latitudes, but the uncertainties in those
data are large.

The above simulation is presented solely as a crude consistency check of the
proposed model. An obvious problem with this approach is that the model does
not extend far enough along the major axis to test the velocity observations
well away from  the photometric center  of \sgg; this  is because the simple
analytical model  constructed in \S5.2 is  spherical, fit  to the minor axis
profile, while  the  kinematic and  distance information  are fit along  the
major axis.

\subsection{Dynamical Implications for the Milky Way Galaxy}

Only the  {\it central\/} density of the  background dark halo  of the Milky
Way (\markcite{mer92}Merrifield  1992) equals the mean  density that we have
calculated for \sgr, so it is not likely that \sgr\ will be tidally shredded
by the Galactic halo acting alone; the disk contributes an important part of
the potential (but  the structure  of   the disk  is unimportant, see   also
\markcite{vel95}Velasquez \& White 1995). The mean density of \sgr\ obtained
above  is   perhaps   10  times   that  of     the  disk at     15~kpc   ---
\markcite{kui89}Kuijken  \&  Gilmore (1989) find that   the disk has density
$\rho_{disk}(15\, {\rm kpc})  \sim 8 \times  10^{-3} \msun ~ {\rm pc}^{-3}$,
while the heavier disk modeled by \markcite{vel95}Velasquez \& White (1995),
which has a total surface density at the solar neighborhood of $72.6 \msun ~
{\rm  pc}^{-2}$, has twice  this, $\rho_{disk}(15\,   {\rm kpc}) \sim  0.014
\msun  ~ {\rm  pc}^{-3}$. The relatively  high density  of  \sgr\ raises the
possibility that  \sgr\ could indeed impart some  damage to the Milky Way as
it is accreted, if it could couple to the disk.

Following  \markcite{ost90}Ostriker (1990), the   absolute maximum amount of
energy that could be imparted to the disk is the total orbital energy of the
satellite, giving an increase of random energy of disk stars of
\begin{equation} 
\Delta v^2 \sim v^2_{orbit} M_{Sgr}/ M_{disk},
\end{equation}
which is 
\begin{equation} 
\Delta v^2 \sim 250^2\times 10^{-2} (M_{Sgr}/  10^9)
(5 \times 10^{10}/ M_{disk}) \sim (45 {\rm km/s})^2.
\end{equation}

If {\it all\/}  of this energy  could be put  into vertical heating, and the
initial disk   has  $\sigma_{z} \sim   20$~km/s  dispersion,  this would  be
increased to $\sim 50$~km/s.  Of course this is  a highly unlikely event ---
the energy absorbed by  the disk would  be spread among its internal degrees
of freedom,  even    neglecting the internal   degrees   of freedom of   the
satellite,    and  of     the      dark halo  and   the     dissipation   of
gas. \markcite{bin92}Binney (1992) has reviewed  the role that accretion may
play in the generation and sustaining  of warps in disks.   The mass that we
have derived  for  \sgr\ is sufficiently   high  to suggest   a role in  the
generation of the Milky Way  warp, as suggested by \markcite{lin96}Lin  {\it
et al.} (1996).

\section{The Properties of \sgr}

\subsection{The Mass/Luminosity Ratio of \sgr}

The volume mass density  we  have derived  above may  be transformed into  a
surface mass density using our adopted shape for \sgr, as:
\begin{equation}
\Sigma \sim \rho \times 2 R_{halo, \,Sgr} \sim 0.03 \times 2 \times
2 \times 10^3 \msun ~ {\rm pc}^{-2} \sim 150 \msun ~ {\rm pc}^{-2}.
\end{equation}
\markcite{mat95a}Mateo {\it et al.} (1995a) estimate that the central
surface brightness of \sgr\ is $\sim \mu_V \sim 25.4$ mag/square
arcsec, which corresponds to $\sim 3 L_\odot ~ {\rm pc}^{-2}.$
Thus an estimate of the central M/L$_V$ of \sgr\ is $\Sigma/\mu_V \sim 50$.

The Gaussian star-count profile derived above contains a total luminosity
\begin{equation}
L_V = \pi I_0 a^2 \sim 3.4 \times 10^6 L_{V\odot}
\end{equation} 
with the  parameter values obtained earlier.  Adopting  all of the parameter
values from the same model fit implies a global $(M/L)_V \sim 300$. However,
model-independent estimates of  the total luminosity  of \sgg\  suggest $L_V
\simgt 10^7 L_{V\odot}$, giving a global $(M/L)_V \sim 100$.

Previous  determinations        of    M/L    ratios    for       dSph   (\eg
\markcite{har94?}Hargreaves, Gilmore, Irwin  \& Carter 1994)  have generally
used King model fits and the expression
\begin{equation} 
{\rho_o \over {\cal L}_o} = \eta {333 \sigma_o^2 \over R_{HB}
I_o},
\end{equation} 
for the central M/L ratio in solar units, where ${\cal L}$ is the luminosity
density and $\eta$ is a parameter close to unity  whose value depends on the
King model fit. This expression gives
\begin{equation} 
{\rho_o \over {\cal L}_{o,V}} = \eta \, 22.2,
\end{equation}
in reasonable agreement with the value obtained above.

\subsection{Dwarf Spheroidal Masses and Chemical Evolution}

The model proposed above results in a central $M/L_V$ for \sgr\ of around 50
in  solar units.  Irrespective  of any detailed   modeling, the high stellar
velocity  dispersion, $\sim 11\kms$, and the  low central surface brightness
noted above, provide incontrovertible evidence that \sgr\ has a high central
mass to light ratio.   Thus \sgr\ joins  the ranks of  those dSph for  which
there  is strong  evidence for substantial   amounts of dark matter,  namely
Draco, Ursa Minor, Sextans, Carina and Leo II.  The most recent estimates of
central mass-to-light ratios are discussed  by \markcite{irw95}Irwin \& \des
(1995), presented in their Table~10. Fornax, Leo I and Sculptor have derived
M/L  values  which remain   (marginally) consistent  with  a normal  stellar
population.

Correlations   between M/L and  other  properties  of  the  dSph potentially
constrain the nature of the dark matter, though in a manner which has yet to
be clarified.   The     then-extant inverse correlations   between  M/L  and
luminosity,  and   M/L  and       mean   [Fe/H], were    interpreted      by
\markcite{dek86}Dekel \&  Silk (1986) as supporting  CDM dark halos, and by
\markcite{lar87}Larson (1987) as  indicative of IMF  variations and baryonic
halos. In  any case,  \sgr\ provides a  striking counter-example  to  these
correlations.

\markcite{lee95}Lee (1995) has  recently presented the available metallicity
---   surface-brightness and  metallicity    --- luminosity data  for  dwarf
galaxies in the Local  Group (his Figure~9).  If  \sgr\ were  to lie on  the
relations for dSphs, then with  a metallicity of $\sim  -1$~dex, it would be
expected to  have a luminosity  $M_V \sim -16$, or  $L_V \sim  2 \times 10^8
L_\odot$, and a  central  surface brightness in  the  V-band of $\simgt  20$
mag/square  arcsec.  The luminosity estimated from  the number of horizontal
branch stars  within the contours of  \IGI\ is $M_V =  -13$, and the central
surface brightness is 25.4  mag/square  arcsec. The Sagittarius  dwarf  thus
lies 3  mag   off  the  luminosity relation,   and 5.5  magnitudes   off the
surface-brightness relation.

The considerable spread in both the age distribution  and the metallicity of
the stars in  \sgg, and especially  of the globular clusters, constrain  the
star formation   history,  and  the chemical   evolution.  While  it remains
possible in principle that Sgr accreted metal-enriched gas repeatedly during
its evolution, the simplest interpretation  is that \sgr\ managed to recycle
gas  through successive generations of  stars, but deciding whether this was
by retention   of chemically-enriched gas,   by  re-capture of  gas  after a
supernovae-driven wind (\cf  \markcite{sil87}Silk, Wyse \& Shields 1987) or
by some other  process, requires  detailed age  and chemical  element  ratio
distributions. Retention of  gas would   suggest either  that Sgr is/was   a
fairly massive  galaxy, with a large  enough escape velocity, or  that there
was  always   a low   star-formation rate, since     the combination of high
star-formation rate and low  escape velocity would  most probably lead to  a
wind (\eg  \markcite{lar74}Larson  1974; \markcite{wys85}Wyse \& Silk 1985;
\markcite{dek86}Dekel \& Silk 1986).

Element  ratio data will   provide unique constraints on  the star-formation
history and gas flows;  a system which forms  stars in bursts separated by a
hiatus, and retains gas, will enrich the gas in iron from Type Ia supernovae
between bursts (\markcite{gil91}Gilmore \&  Wyse 1991), allowing stars  with
very low oxygen-to-iron ratios to form in the subsequent burst. By contrast,
gas accreted from the  inter-stellar medium of  the proto-Milky Way, or  gas
`lost' in  a temporary  wind  from Sgr   and  re-accreted, mixed with   extra
Galactic gas, should have the  high oxygen-to-iron ratio evident in Galactic
field halo stars of the age of the Sgr clusters.

The escape  velocity   from \sgr\ is  the  relevant  quantity for winds  and
chemical  evolution considerations and  may  be  evaluated with the  present
model for the structure of \sgr. An homogeneous sphere, truncated at $r=a $,
has an escape velocity from $r < a$ of:
\begin{equation} 
{1 \over 2} v_e^2 = 2 \pi G \rho (a^2 - {1 \over 3} r^2).
\end{equation} 
For  $a  =  1.2$~kpc,  this  gives $v_e(r=0)  \sim  90$~km/s. This  value is
sufficiently high that \sgr\ may be expected to retain a substantial mass of
gas, even while forming stars fairly actively.

For  comparison, the  escape  velocity from  a  self-gravitating system with
power-law density profile  $\rho(r) \propto r^{-\alpha};  2 < \alpha < 3$ is
$v_{esc} = \sqrt{2} ~ v_{circ} /\sqrt{(\alpha -2)}$, where $v_{circ}$ is the
circular   velocity parameterizing  the  gravitational potential.   Assuming
approximately     isothermal     velocity   dispersions,    $v_{circ}\approx
\sqrt{\alpha} ~  \sigma_{1-D}$,  with $\sigma_{1-D}$  the observed   central
stellar  velocity  dispersion. Thus,   for a  system  with power-law density
profile exponent  $\sim  3$, and $\sigma=10$~km/s,  the  escape  velocity is
$v_{esc} \approx \sqrt{6}.\sigma_{1-D} \approx 25\kms$.

Thus, while  this is at best  a partial explanation, a constant-density mass
profile  of the form   adopted here  does reduce  somewhat  the  discrepancy
between calculated escape  velocities  from stellar systems  which evidently
were  able to   form chemically-enriched  stars,  and  that escape  velocity
required to  prevent  supernova-driven  winds,  which  would  prevent  later
generations of stars from being formed.

\section{Summary}

We present a  self-consistent picture of the  structure of \sgr, the nearest
satellite galaxy to  the Milky Way. We combine  extant data on its distance,
metallicity,  age   and  stellar populations    with  new  photometric   and
spectroscopic data which determine its size, shape and dynamics.

The geometrical picture derived  from these data is that  \sgg\ is a prolate
body with axis ratios $\sim$~3:1:1.  The center of Sgr is $\sim 25$~kpc from
the  Sun and $16 \pm  2 \kpc$ from the Galactic   center.  \Sgr\ is oriented
approximately  perpendicular to  the  plane   of  the Galaxy  from  Galactic
latitude $b=-4^\circ$  to   $b=-26^\circ$.   Its longest   dimension extends
$\simgt 9 \kpc$ along the coordinate line $l=5^\circ$.  \Sgr\ contains a mix
of stellar  populations ranging from relatively old  stars --- many RR~Lyrae
stars are observed and at least one of its  four globular clusters is as old
as the oldest Galactic   halo clusters ---   to intermediate age  stars  ---
several Carbon stars have  been identified. The dominant population however,
is  10 to 14  Gyr  old and has  corresponding mean  abundances between ${\rm
[Fe/H]=-0.8}$ and ${\rm   [Fe/H]=-1.2}$. The full abundance  range  observed
covers $\simgt 1$dex around this mean.

The kinematic data presented  include a first study  of the proper motion of
Sgr, together with intermediate ($\sim 12 \kms$) and  high ($\sim 2.5 \kms$)
accuracy radial velocity measurements of its member stars.  Due to a lack of
known extra-galactic  reference  objects in   these low   Galactic  latitude
fields,  the proper motion was  measured with respect to Galactic foreground
stars.  This  approach    means that only  the    motion  in  the  direction
perpendicular to   the  plane of  the   Galactic disk is   constrained; this
component of    its velocity is  $250 \pm   90 \kms$,  directed  towards the
Galactic Plane.

The radial velocity data determine the kinematics at several positions along
the major axis from $b=-26^\circ$ to  $b=-12.5^\circ$.  These data show that
\sgg\ has an internal velocity dispersion, of  $11.4 \pm 0.7 \kms$, which is
formally consistent with  being  constant over the  face of  the galaxy. The
mean velocity   gradient $dv/db$ is  small  in the  central regions  but the
amplitude increases to $dv/db=-3 \kmsd$ over the outermost three degrees for
which we have data.

These kinematic data indicate that Sgr has no significant rotation about its
minor axis, but they do not constrain the rotation about its major axis, due
a lack of sampling  in the latitude  range $b>-12.5^\circ$. This uncertainty
in the   major  axis rotation   rate is  a   source of   uncertainty in  the
determination of  the orbit  of   Sgr. However, for all  plausible  internal
rotation  rates the orbital  period is  $\lta 1 \Gyr$  and the  perigalactic
distance is $\sim 12 \kpc$. This orbital period is  much shorter, by about a
factor   of  $\simgt 10$,  than  the    age  of  the   bulk of  its  stellar
population.   We  argue  that   dynamical  friction will   not have  induced
significant   orbital decay, and that it   is  unlikely that  \sgr\ has been
captured  recently,  so  it   has apparently   survived  many perigalacticon
passages.

Given these  observational constraints, the  consistency of models of Sgr is
explored  in  the context of   recent results from  numerical simulations of
disrupting satellite galaxies.    Perhaps the most  restrictive result  from
these  simulations  is that tidally  disrupted  debris  retains the velocity
dispersion of its progenitor. Thus,   the measured velocity dispersion is  a
clear   indication  that  \sgr\    did   not  have  a  much    more  massive
progenitor. Furthermore,  if it had  been significantly more  massive in the
past, we would expect to find its `missing mass' as a substantial population
of Sagittarius dwarf debris  --- globular clusters  and stars --- along  its
dispersion orbit; however, this is not observed.

The  most conservative  assumption  is to  adopt  a  mass distribution  that
follows the distribution of luminous matter. \Sgr\  would then approximate a
King-model; yet King-models that satisfy  the radial light distribution  and
central  velocity dispersion    lead  to substantial  mass    loss   at each
perigalacticon   passage    (\markcite{vel95}Velazquez \&      White,  1995;
\markcite{joh95}Johnston \etal  1995), inconsistent with  survival until the
present time. This   simple mass-follows-light  model has  too   low a  mass
density at the photometric edge to  inhibit tidal dissolution over $\sim 10$
perigalacticon passages.   Sagittarius  and,  by  implication,  other  dwarf
spheroidal galaxies cannot have a mass distribution in which the dark matter
profile is similar to that of the luminosity.

We therefore investigate the  next simplest model:   one that minimizes  the
total mass required for survival. Such a model  has a mass distribution with
a large core radius (as is observed in gas-rich dwarf galaxies), so that the
mass density is approximately  constant  over the dwarf. Consequently,   the
mass to light ratio increases radially outwards from the center of Sgr. This
model yields a total $M/L \sim 100$ and  a total mass  of $\sim 10^9 \msun$.
This  model is  consistent with   only a  small   amount  of previous  tidal
dissolution --- so \sgr\  can survive until the  present day, is  consistent
with the observed central velocity dispersion and implies that its orbit has
not decayed significantly over its lifetime.

An  additional feature of this  model is that  the  escape velocity from the
center of  the  potential  well is significantly  higher   than it is   in a
power-law  model.  This reduces to some  extent  a  difficulty with standard
models of the chemical  evolution of dSph  galaxies, that winds generated by
star formation and chemical  enrichment should remove their  gas, preventing
chemical self-enrichment.

Contrary to previous estimations, this model suggests that \sgr\ will resist
tidal disruption by  the Milky Way  and remain an internally bound satellite
galaxy  until dynamical   friction significantly   reduces its  perigalactic
distance,  on  a timescale of  a  Hubble time.   The robustness of  \sgr\ is
consistent with  other determinations of  the contribution of disrupted dSph
to  the field stellar  halo,   which have  found   this to  be  $\sim  10\%$
(\eg \markcite{una96}Unavane, Wyse \& Gilmore 1996).

\bigskip

Our collaboration was  supported by the  NSF (INT-9113306) and by NATO.  RAI
expresses gratitude  to the Killam  Foundation (Canada)  and to  the  Fullam
Award for support. RFGW  acknowledges the Seaver  Foundation, and thanks Jay
Gallagher  for  many discussions  about    dwarf galaxies.  The   Center for
Particle Astrophysics is funded by the NSF.  \vfill\eject

\clearpage

\figcaption[Ibata.fig1.ps]{Isopleth  contours of  the  tip  of the  Sgr main
sequence, derived from APM measurements of UKST survey plates to the East of
the center of  Sgr, are shown superimposed on  the discovery isopleth map of
the red clump stars  from IGI 1994. The positions  of the  globular clusters
potentially associated with the Sgr dwarf: M54, Ter~7,  Ter~8, and Arp~2 are
also indicated, as  is the position of  the foreground globular cluster M55.
Contours for the Sgr main sequence tip start at $\approx$1 arcmin$^{-2}$ and
increment by $\approx$5 arcmin$^{-2}$.}

\figcaption[Ibata.fig2.ps]{The magnitude distribution  of stars in the color
strip $1.05 < V-I < 1.17$ in a field near the center  of \sgr\ (the data are
from the  comparison field of Sarajedini \&  Layden 1995). The  narrow local
maximum at $V=18.25$  corresponds to the red  clump population in \sgr.  The
fitted model described in the text is superimposed.}

\figcaption[Ibata.fig3.ps]{The color-magnitude  diagram  of  a  $7'\times7'$
field  at  ($\ell=15.48^\circ,b=-15.09^\circ$),   close  to   the center  of
\sgg. Galactic disk dwarfs predominate in the blue vertical feature at ${\rm
V-I \sim 0.8}$, ${\rm V < 17.5}$, while Galactic bulge subgiants predominate
in the intermediate color vertical feature at ${\rm V-I  \sim 1.1}$, ${\rm V
< 17.5}$. The tip of \sgg\ main sequence is  seen near ${\rm V-I \sim 0.6}$,
${\rm  V \sim  21}$,   while its red clump    is at ${\rm  V-I=1.1}$,  ${\rm
V=18}$.  The sloping feature   from ${\rm V-I=1.3}$,  ${\rm  V=18}$ to ${\rm
V-I=1.5}$, ${\rm V=16}$ is the RGB of \sgr.}

\figcaption[Ibata.fig4.ps]{The heliocentric  distance  to 26 fields  along a
portion of  the  major axis  of \sgg\ is   shown as a  function  of Galactic
latitude. As  described  in  the  text, a correction   for the  variation in
extinction was  applied to the photometry,  and  distances were derived from
the fitted V magnitude of the red clump in each field. The  dotted line is a
$\chi^2$ fit to these data (reduced $\chi^2=1.17$).}

\figcaption[Ibata.fig5.ps]{The  stars  whose radial  velocity   measurements
confirm that they are   members of  \sgr\  are  plotted onto  the  isopleths
presented in    Figure~1.  The plusses  represent  the   high precision CTIO
measurements, while the filled circles correspond to the AAT measurements. A
grid of Galactic coordinates has been superimposed.}

\figcaption[Ibata.fig6.ps]{Radial velocities for stars  which are members of
\sgg, corrected for the  Galactic rotation  of  the Local Standard  of Rest,
plotted   as a function   of Galactic latitude. The   velocities of the four
globular clusters of the Sgr system are indicated, with  the latitude of the
globular cluster being at the center of the relevant marker.}

\figcaption[Ibata.fig7.ps]{The velocity data  are subdivided into the fields
shown  in the central  panel.  For convenience, the   isopleth map of  \sgg\
derived by \IGI\ and Galactic coordinate  lines have been superimposed.  The
CTIO velocity data for  fields (from left to  right) f1, f5,  f6 and  f7 are
displayed in the  form of histograms  on the top  row of this diagram, while
the AAT data in all eight fields are displayed on the bottom two rows of the
diagram. In all of the  histograms displayed  on  this page, the  horizontal
coordinate is a Galactocentric radial velocity in km/s and the vertical axis
is the number of stars per bin. A fitted Gaussian to each of these data sets
has been overlaid and the fitted mean (and dispersion  for the CTIO data) is
indicated.}

\figcaption[Ibata.fig8.ps]{The  run of  mean  Galactocentric radial velocity
along the major axis of \sgr\ (assumed to be  parallel to a line of constant
Galactic longitude)  is shown in the left  hand panel.  The right hand panel
presents the velocity dispersion profile along the major axis.}

\figcaption[Ibata.fig9.ps]{Limits on minor axis rotation  in \sgg. The three
panels   show different Galactic latitude slices   through the velocity data
presented  in  Tables~2.A,~2.B  and   Ibata    \&  Gilmore   (1995a),  their
Table~B1. Straight-line fits provide no evidence for  a gradient in velocity
as a function of Galactic longitude in  these three fields.The adopted major
axis is aligned with the Galactic coordinate line $\ell=5^\circ$.}

\figcaption[Ibata.fig10.ps]{The orbit of a test particle which best fits the
kinematic data of \sgg\ in the Milky Way  potential described in the text is
shown projected  on the   available data.  The   left hand panel  shows  the
projected velocity of the orbit along the lines of sight to the fields which
determine the fit. Formally, this orbit is an  acceptable description of the
data, despite the systematic difference between  the model and the data. The
right hand panel compares the heliocentric distance of the guiding center of
the  orbit   fit to  the  velocity data  to   the heliocentric distance data
presented  in Figure~4.  The modeled orbit  is an  acceptable fit to  these
distance constraints (reduced $\chi^2=1.8$).}

\figcaption[Ibata.fig11.ps]{The orbit of \sgg, integrated  over $1 \Gyr$, is
shown in  the $x$ -- $z$   plane. The `star'   symbol represents the present
position of the Sun, while   the open ellipse  (drawn  to scale) gives   the
present position of \sgr. The radial period of this orbit is $0.76 \Gyr$.}

\figcaption[Ibata.fig12.ps]{The assumptions used in \S3  to derive the orbit
of \sgg\ are checked for consistency against  the analytical model (given in
\S5.2), which proposes that the mass distribution of \sgg\ is dominated by a
spherical  dark matter (DM)  halo.  The solid  lines  show (on the left-hand
panel)  the radial velocity  and (on the  right-hand panel) the heliocentric
distance along the orbit of the center of mass (derived in \S3). The stellar
component  of  \sgg\ is   modeled as an    isothermal population of  $10^4$
non-interacting    tracer particles which move  under   the influence of the
(fixed) Galactic potential  plus the potential of  the Sgr DM halo (which is
assumed to move on the orbit of the center of mass). The model is integrated
for 10~Gyr and stopped at the present position of \sgg. The tracer particles
are then  binned along the  major axis in groups  of 200; the mean projected
radial velocity  in  these bins is  plotted  (dots) in the left-hand  panel,
while mean heliocentric  distance is plotted  in the right-hand  panel, both
quantities  as a function of  Galactic latitude.  Since the projected radial
velocity and  distance  profiles  to   the modeled  stellar  component  are
approximately equal  to the radial velocity and  distance along the orbit of
the center of mass, the assumptions stated in \S3 hold.}

\clearpage

{\sc TABLE} 1. Physical parameters of the globular clusters of \sgr.

{\sc TABLE} 2.A. Radial velocity data from 1994 AAT run.

{\sc TABLE} 2.B. Radial velocity data from 1994 CTIO run.

{\sc TABLE} 3. Mean velocities and velocity dispersions in the
observed fields.

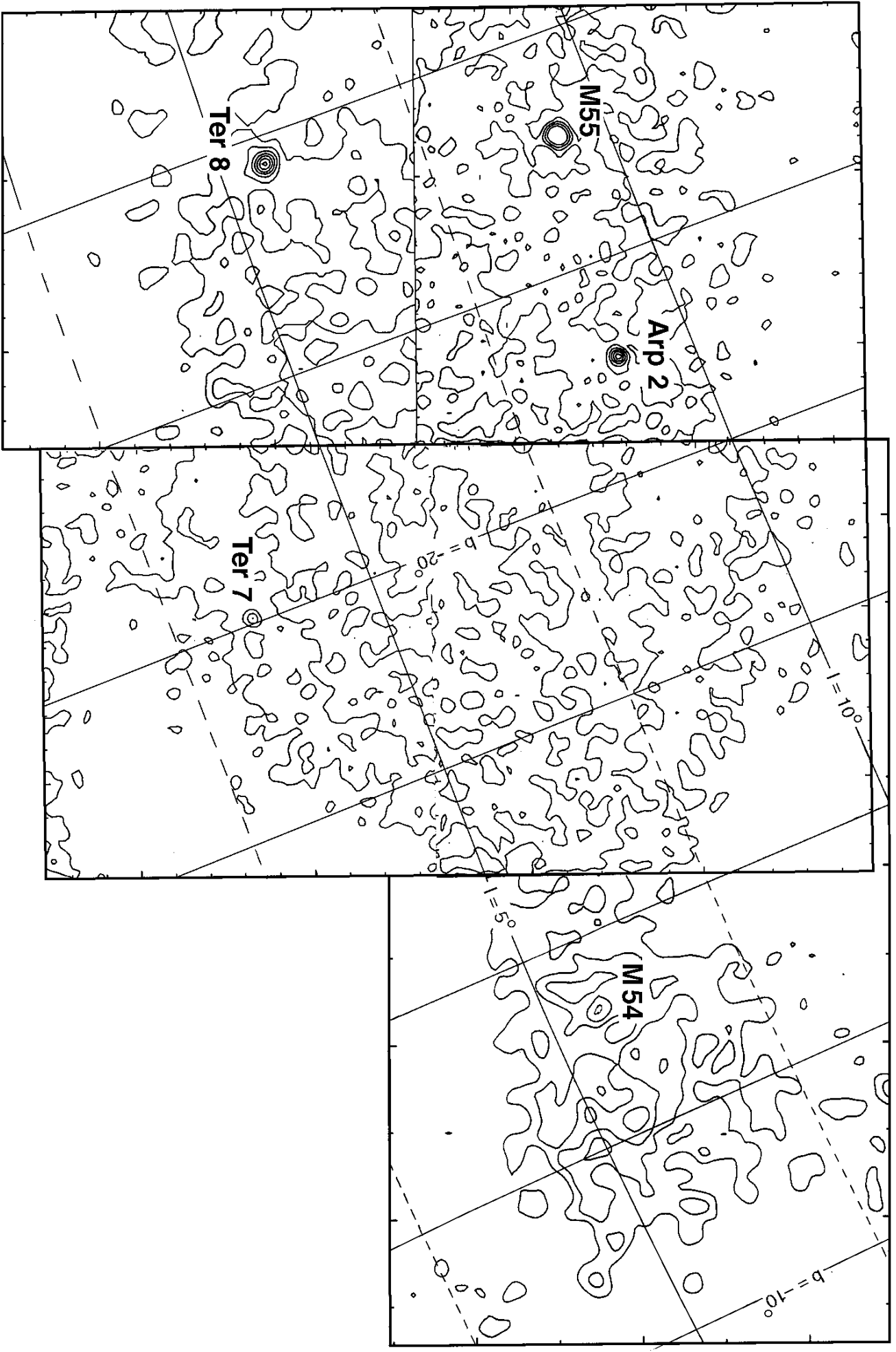
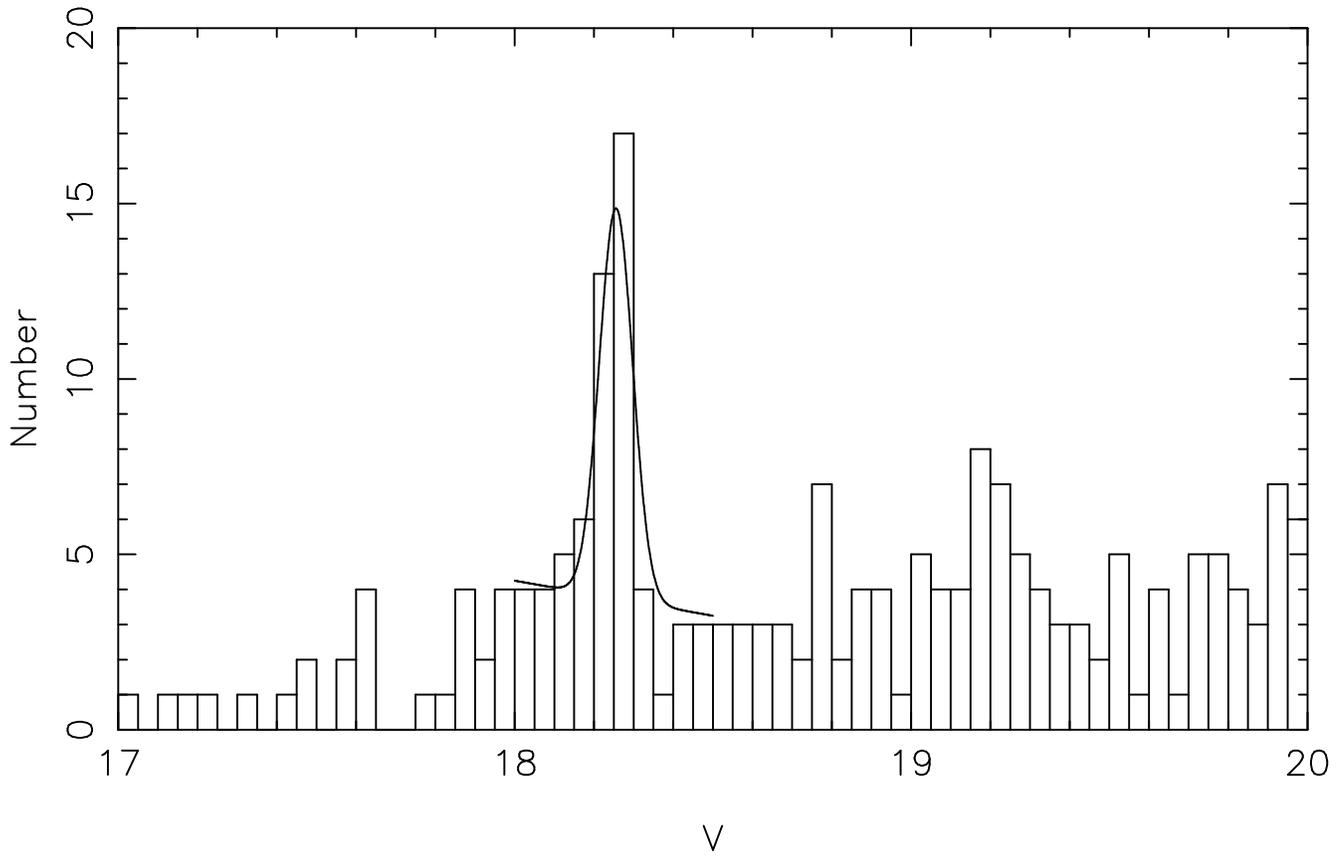
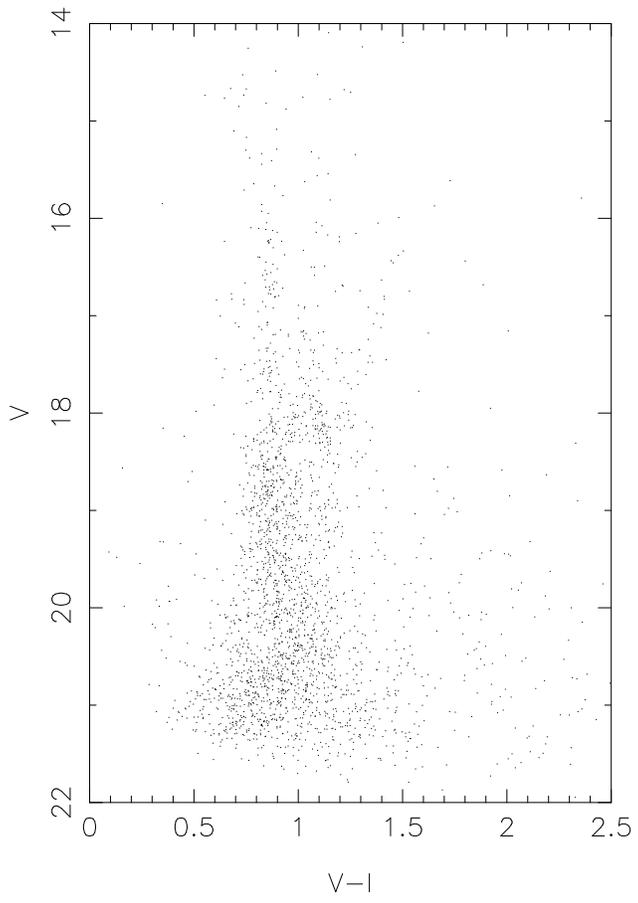
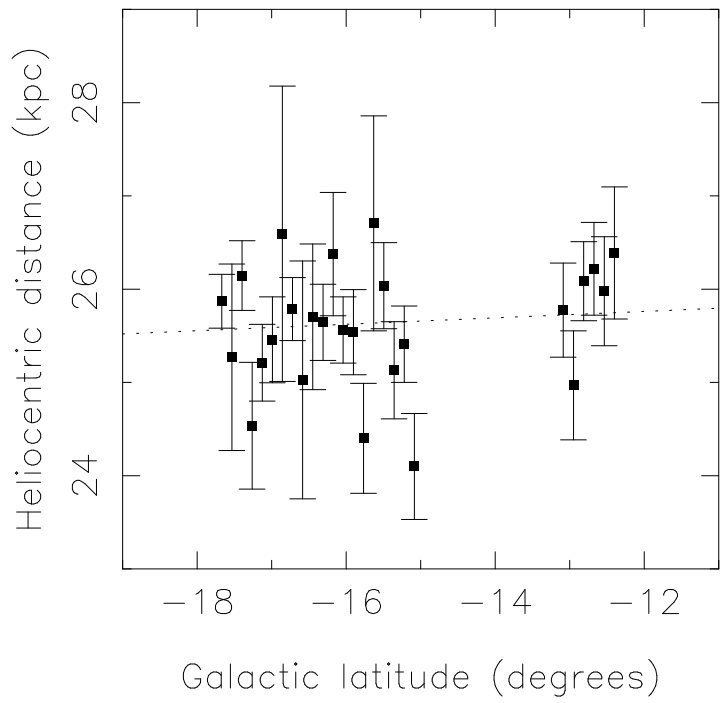
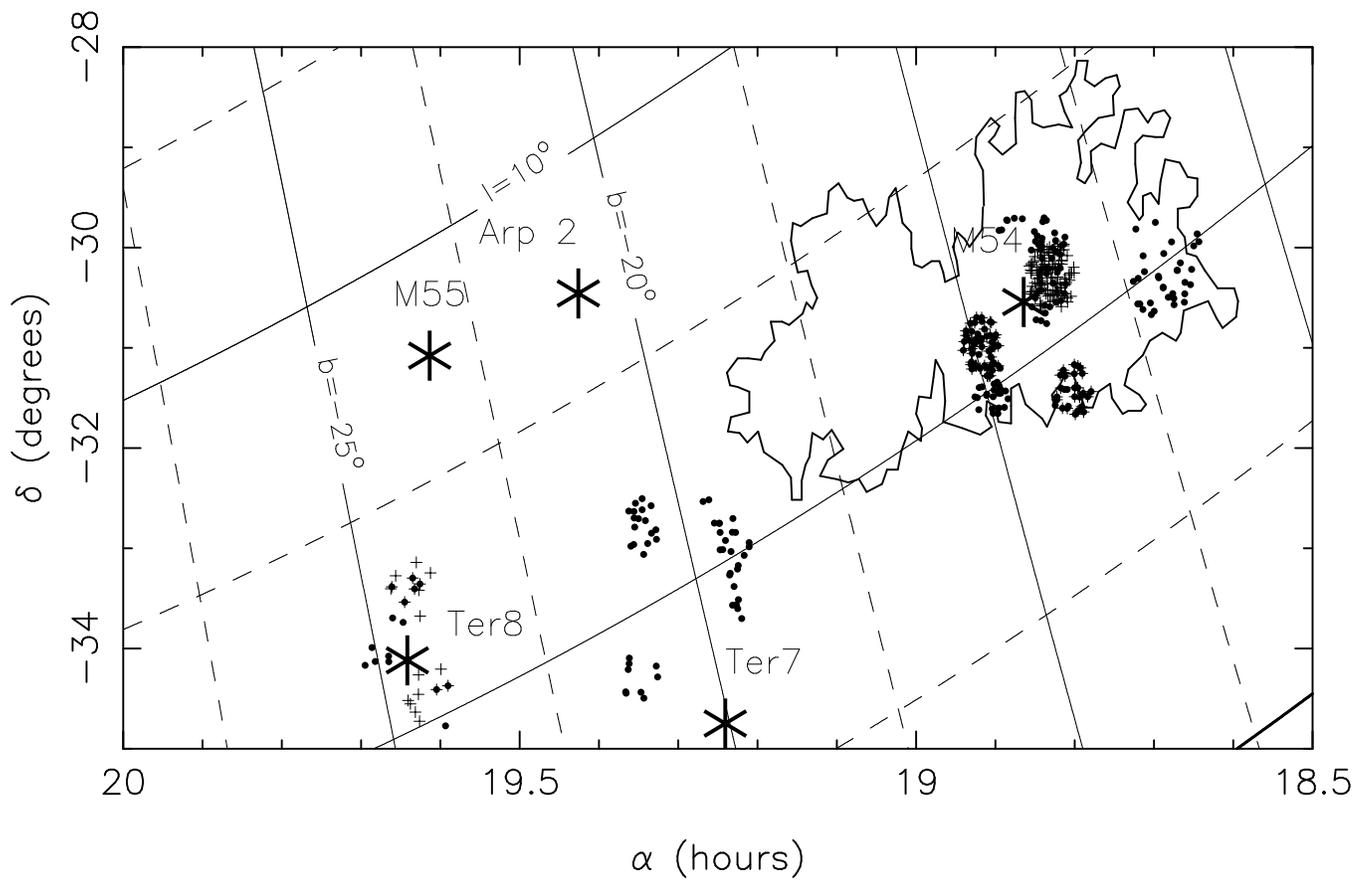
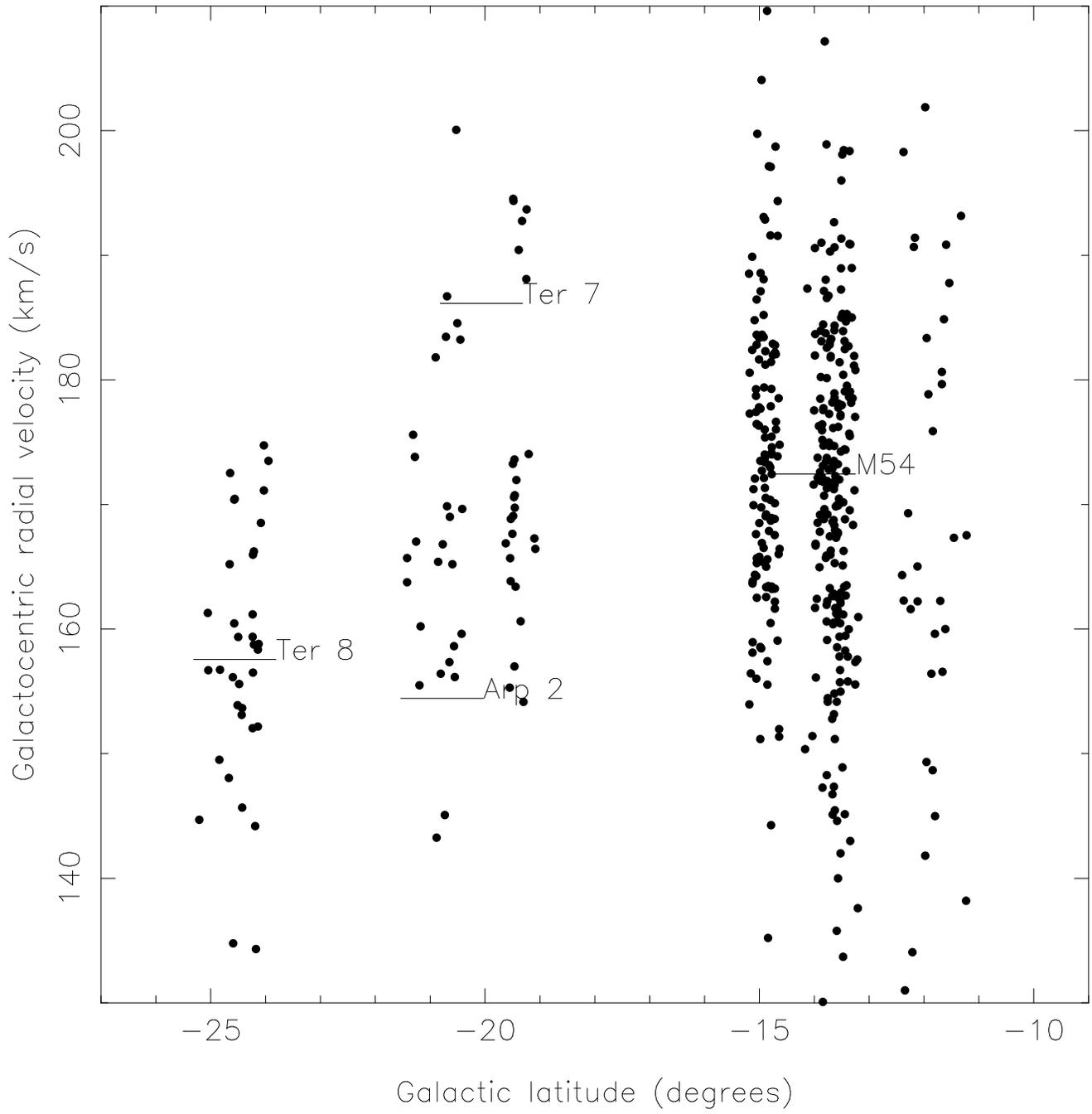
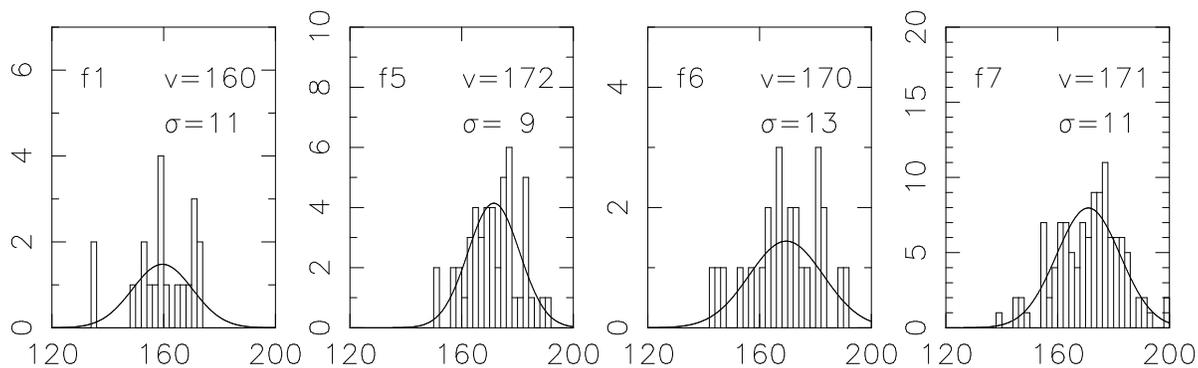
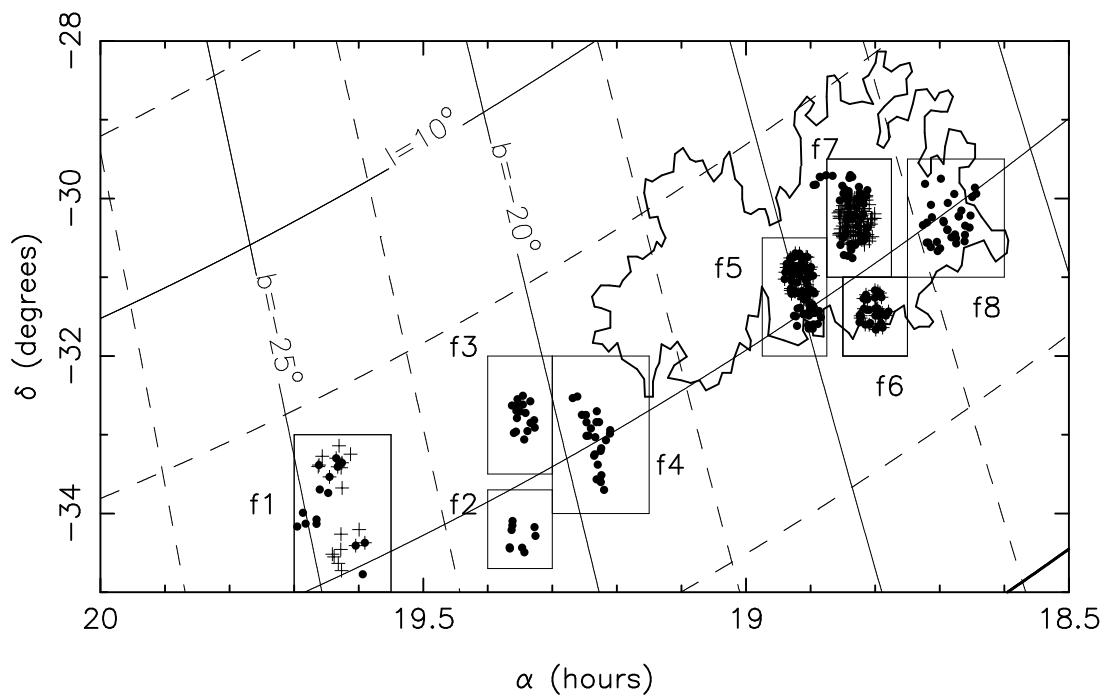
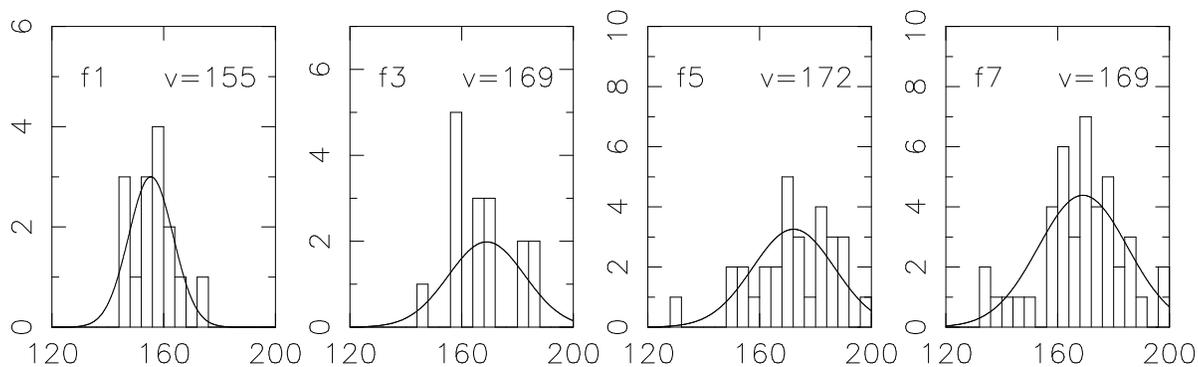
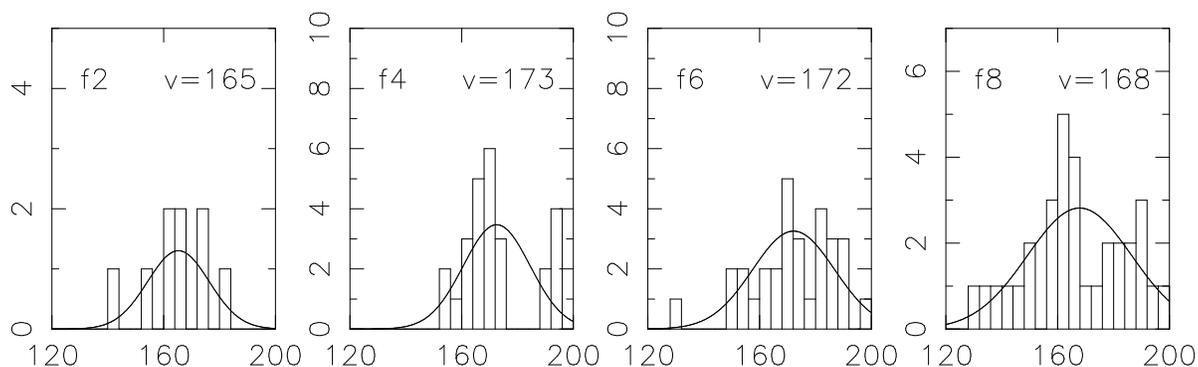
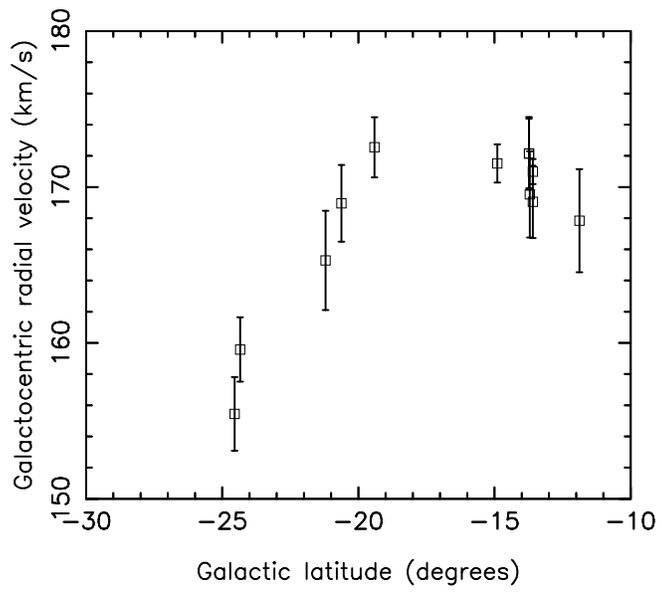
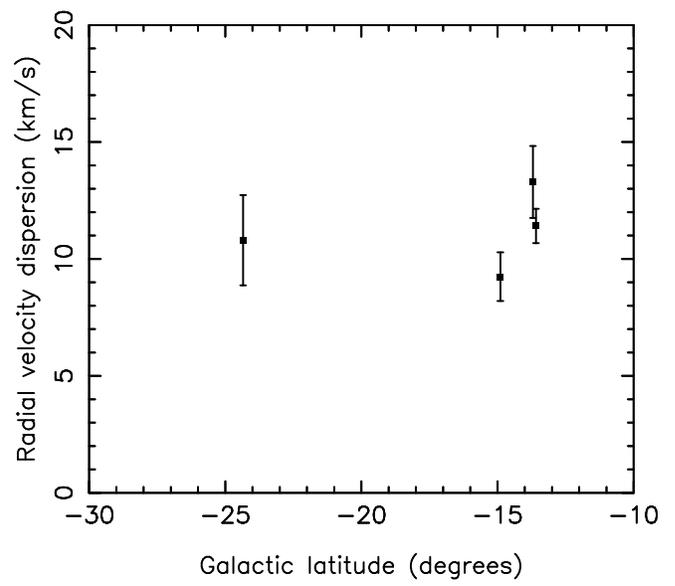
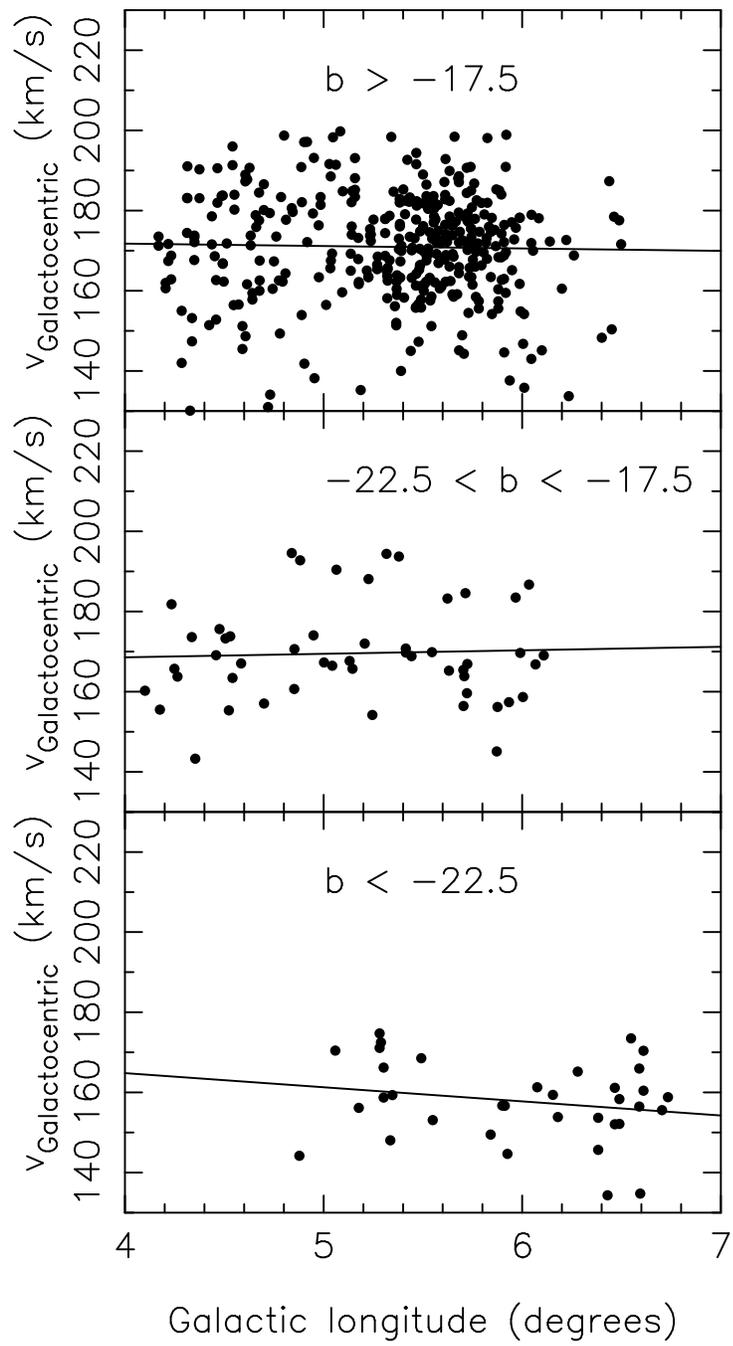
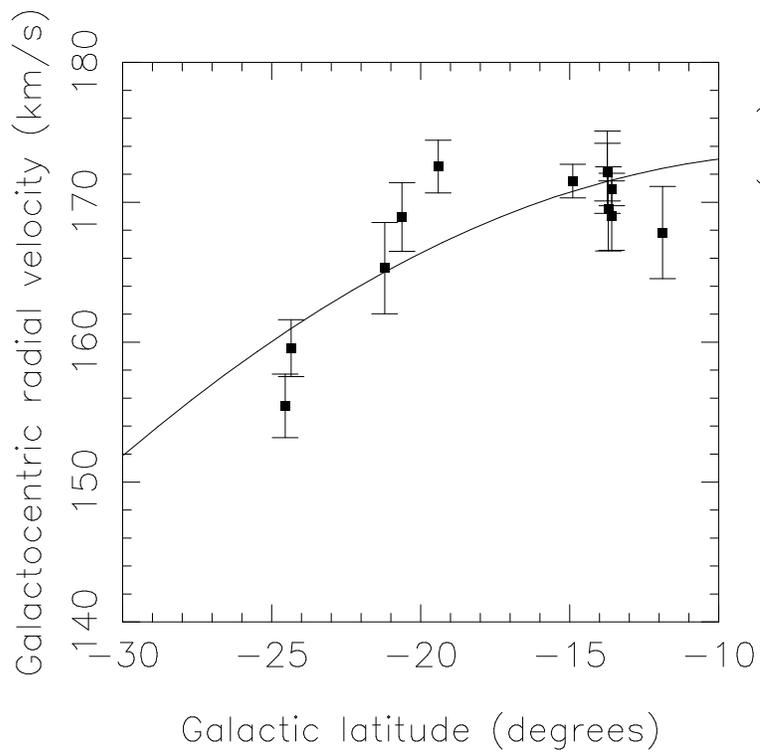
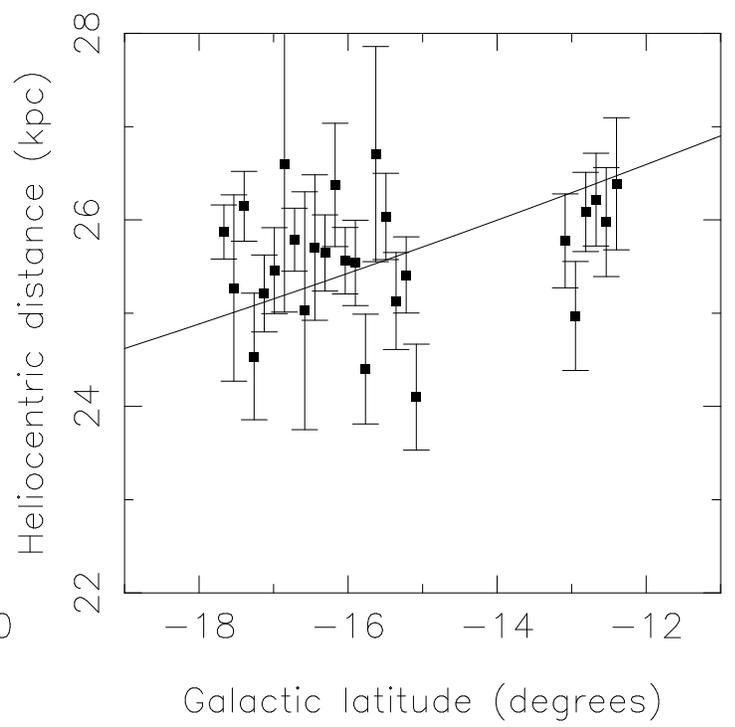
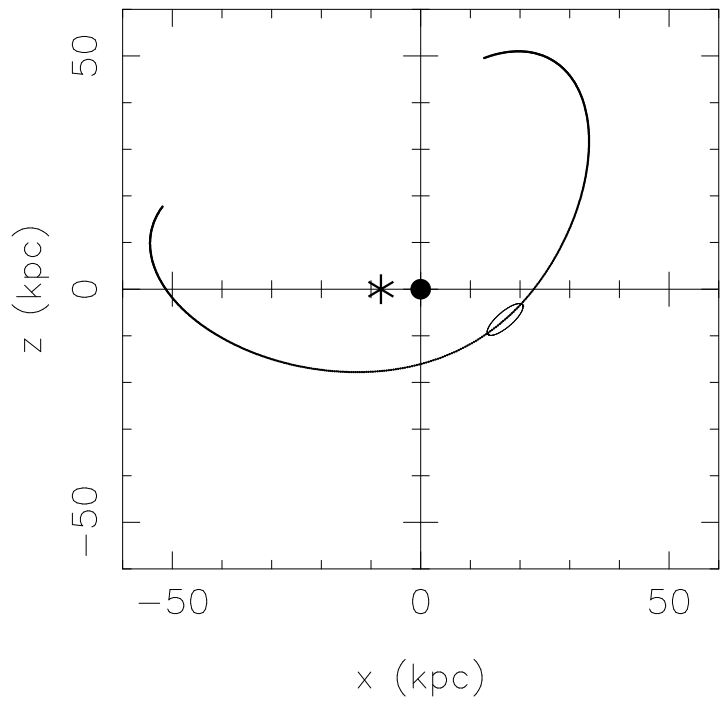
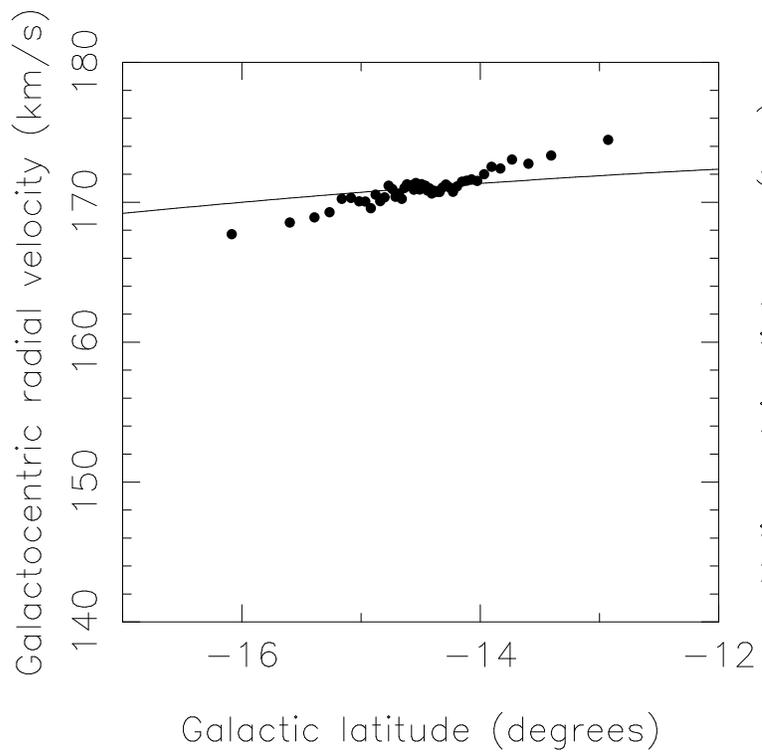
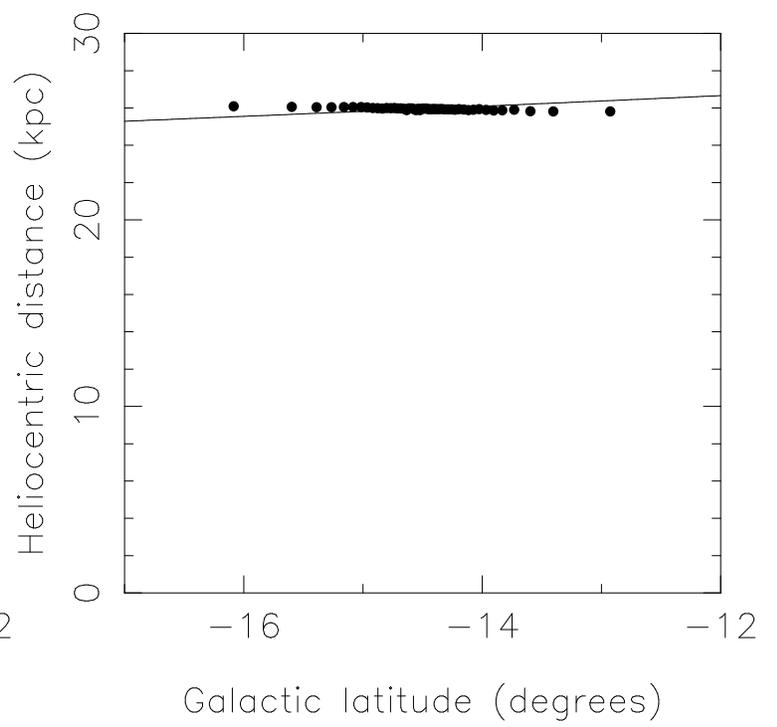
\end{document}